\documentclass[superscriptaddress,twocolumn,pre]{revtex4}
\usepackage{amsmath}
\usepackage{amssymb}
\usepackage{graphicx}
\usepackage{hyperref}
\usepackage{amsfonts}
\usepackage{bbm}
\usepackage{doi}

\newcommand{\beginsupplement}{%
        \setcounter{table}{0}
        \renewcommand{\thetable}{S\arabic{table}}%
        \setcounter{figure}{0}
        \renewcommand{\thefigure}{S\arabic{figure}}%
        \setcounter{equation}{0}
        \renewcommand{\theequation}{S\arabic{equation}}%
        \setcounter{subsection}{0}
        \renewcommand{\thesubsection}{\Roman{subsection}}%
      }

\makeatletter
\newcommand*{\rom}[1]{\expandafter\@slowromancap\romannumeral #1@}
\makeatother

\newcommand{\nocontentsline}[3]{}
\newcommand{\tocless}[2]{\bgroup\let\addcontentsline=\nocontentsline#1{#2}\egroup}

\begin{document}
\title{Geometry of Bend: Singular Lines and Defects in Twist-Bend Nematics}
\author{Jack Binysh}
\thanks{These authors contributed equally to this work.}
\affiliation{Mathematics Institute, Zeeman Building, University of Warwick, Coventry, CV4 7AL, United Kingdom.}
\affiliation{Department of Physics, University of Bath, Claverton Down, Bath, BA2 7AY, United Kingdom.}
\author{Joseph Pollard}
\thanks{These authors contributed equally to this work.}
\affiliation{Mathematics Institute, Zeeman Building, University of Warwick, Coventry, CV4 7AL, United Kingdom.}
\author{Gareth P. Alexander}
\email{G.P.Alexander@warwick.ac.uk}
\affiliation{Department of Physics and Centre for Complexity Science, University of Warwick, Coventry, CV4 7AL, United Kingdom.}

\begin{abstract}
We describe the geometry of bend distortions in liquid crystals and their fundamental degeneracies, which we call $\beta$ lines. These represent a new class of line-like topological defect in twist-bend nematics where the bend is generically non-zero. We present constructions for smectic-like textures containing screw and edge dislocations, grain boundaries and focal conics, and also for vortex-like structures of double twist and Skyrmions. We analyse their local geometry and global structure, showing that their intersection with any surface is twice the Skyrmion number. Finally, we demonstrate how arbitrary knots and links can be created and describe them in terms of merons, giving a new geometric perspective on the fractionalisation of Skyrmions.
\end{abstract}
\date{\today}
\maketitle

Fresh perspectives invariably accompany the discovery of a new phase: The experimental discovery of the twist-bend nematic phase~\cite{cestari2011,borshch2013,chen2013} invites fresh consideration of nematic geometry and topology. The twist-bend nematic is a fluid mesophase in which the nematic orientation exhibits a heliconical modulation with nanoscale pitch and modest cone angle~\cite{jakli2018}. It occurs in compounds with a bent core architecture (banana molecules) and is characterised by a preferred state of non-zero bend distortion~\cite{dozov2001,shamid2013}. The generic geometrical and topological features of bend distortions are thus a natural vehicle for describing the structural degeneracies and defects of the twist-bend phase, however, they arise more generally and apply to any material or system described (even in part) by a unit vector, or line, field.

Geometric elastic distortions pervade soft matter physics~\cite{kamien2002}, providing a common conceptual framework for understanding many different materials as well as numerous methods -- including boundary conditions, substrate topography and surface curvature -- for designing or controlling properties and functionality~\cite{vitelli2013,tran2016,napoli2012,matsumoto2015,vu2018,ellis2018,white2015,mostajeran2017,aharoni2018}. Geometric methods also relate to topological properties through the Gauss-Bonnet theorem and Berry phase physics, so that geometric degeneracies possess both elastic and topological significance, giving them greater potential for material control. A common feature of many materials are structural degeneracies along lines or curves, with examples including flux lines in superconductors~\cite{abrikosov1957}, fluid vortices~\cite{irvine2018}, nodal lines in optical beams~\cite{dennis2010}, C lines in electromagnetic fields~\cite{nye1983}, defect lines in liquid crystals~\cite{deGennesProst} and umbilic lines in general~\cite{machon2016}. In many instances these lines are fundamental to the organisation and properties of the entire material, simultaneously characterising it and offering a mechanism for controlling and engineering specific responses.

In this Letter, we introduce a new line-like geometric degeneracy associated to zeros of the bend in a unit vector field, that we call $\beta$ lines. These lines occur in all materials with vector or orientational order, such as liquid crystals and ferromagnets, but have added significance when there is an energetic preference for non-zero bend, and in such materials $\beta$ lines are a new type of topological defect. We give constructions of both smectic-like textures and Skyrmion states in twist-bend nematics and characterise them in terms of their $\beta$ lines. We provide a description of the local structure of generic $\beta$ lines and show that their signed intersection number with a surface gives a Skyrmion count. Finally, we show how complex three-dimensional textures encoding knotted $\beta$ lines may be constructed, analogous to the `heliknotons' recently created experimentally in cholesterics~\cite{tai2019}, and characterise them in terms of merons. Additional illustrations of each texture discussed, as well as further technical details and examples of the mathematical constructions introduced, are presented in the Supplemental Material.

Orientational order is described by a unit vector ${\bf n}$, called the director. Nematic symmetry, ${\bf n}\sim -{\bf n}$, corresponds to alignment that is line-like, rather than vectorial. The bend ${\bf b} = ({\bf n} \cdot \nabla){\bf n}=-{\bf n}\times(\nabla\times{\bf n})$ is the curvature of the director integral curves; it is a globally defined vector whose sign does not reverse under ${\bf n} \to -{\bf n}$. As ${\bf n}$ is a unit vector the bend is everywhere orthogonal to it, ${\bf b}\cdot{\bf n}=0$. Thus, although bend is a vector field in three-dimensional space, it is atypical, having only two degrees of freedom. In particular, the set of points where it vanishes --- geometrically, the set of inflectional points in the integral curves of $\bf n$ --- is one-dimensional, and forms a collection of fundamental curves in the material that are characteristic of it; we call them $\beta$ lines.

\begin{figure*}[tb]
\centering
\includegraphics[width=0.99\linewidth]{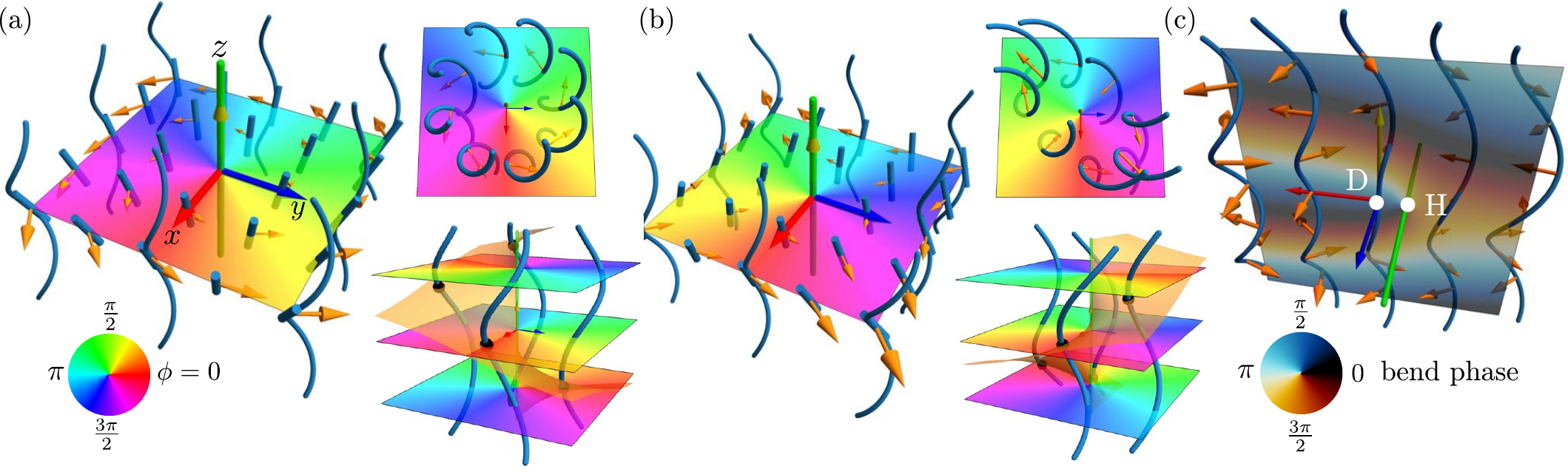}
    \caption{Examples of $\beta$ lines in smectic-like twist-bend singularities. Director and its integral curves shown in blue, with the bend vector in orange and $\beta$ lines in green. (a, b) Screw dislocations in the helical phase of the twist-bend ground state, with strength $s=+1$ (a) and $s=-1$ (b). Colour indicates the phase $\phi$ from \eqref{eq:tb_smectic}. Insets show the winding of the helical phase around the $\beta$ line from multiple perspectives. Orange surfaces in the bottom inset show $\phi=0$ with points of equal phase along the integral curves marked by black dots. (c) Edge dislocation in $\phi$; colour indicates the $xy$ angle of the bend vector. In passing from negative to positive $x$ the director integral curves pick up an extra turn, implying the existence of a $\beta$ line. The $\beta$ line location is not the phase singularity in $\phi$ (D), but the hyperbolic point H where $\nabla\phi$ vanishes.} 
    \label{fig:examples}
\end{figure*}

A model system for exploring the significance of $\beta$ lines is the twist-bend nematic. It may be described by a Frank free energy with negative bend elastic constant~\cite{dozov2001}, or by coupling the bend of the nematic director to a vector polarisation, ${\bf p}$, coming from the `banana' shape of the constituent molecules, with a free energy~\cite{shamid2013}
\begin{equation}
F = \int \frac{K}{2} \bigl| \nabla {\bf n} \bigr|^2 - \lambda {\bf b} \cdot {\bf p} + \frac{C}{2} \bigl| \nabla {\bf p} |^2 + \frac{U}{4} \bigl( 1 - |{\bf p}|^2 \bigr)^2 \, dV ,
\label{eq:F}
\end{equation}
where $K$ is a Frank elastic constant, $\lambda$ is a coupling constant, $C$ is an elastic constant for the polarisation, and $U$ sets the scale of the bulk ordering energy. This favours the heliconical director ${\bf n} = \cos\theta \,{\bf e}_z + \sin\theta ( \cos qz \,{\bf e}_x + \sin qz \,{\bf e}_y )$, with the conical angle $\theta$ and helical wavevector $q$ determined by the elastic moduli~\cite{shamid2013,jakli2018}. The integral curves of the director are helices of constant curvature and torsion; the bend ${\bf b} = q \sin \theta \cos \theta (- \sin qz \,{\bf e}_x + \cos qz \,{\bf e}_y )$ has constant magnitude and rotates at the same rate as the director. We review the geometry of the heliconical director and the free energy \eqref{eq:F} in the Supplemental Material, which includes Refs.~\cite{alexander2018,machon2019,Machon,selinger2019,kats2014,pajak2018}. On scales large compared to the helical pitch ($2\pi/q$) the twist-bend phase has the same elastic energy as a smectic~\cite{kamien1996,parsouzi2016,meyer2016,radzihovsky2011} and exhibits all the features, textures and defects of a bone-fide smectic, despite there being no mass-density wave. These smectic-like defects are all associated with $\beta$ lines; we remark that they are revealed by the director field and the degeneracies of its bend despite many of the textures we consider being nullhomotopic and hence invisible to the traditional homotopy theory methods.

We consider first screw dislocations in the helical integral curves of the twist-bend ground state, Fig.~\ref{fig:examples}. Here, the phase of the helices winds by $2\pi s$ on a circle enclosing the screw axis, where $s$ is the integer dislocation strength; we show $s = +1$ in Fig.~\ref{fig:examples}(a) and $s = -1$ in Fig.~\ref{fig:examples}(b). 
There is the same winding number in the bend (orange arrows), guaranteeing the existence of a $\beta$ line. These textures are captured by the director field
\begin{equation}
{\bf n} = \cos\theta \,{\bf e}_z + \sin\theta \bigl[ \cos\phi \,{\bf e}_x + \sin\phi \,{\bf e}_y \bigr] ,
\label{eq:tb_smectic}
\end{equation}
where $\phi = qz + s \arctan(y/x)$ and the cone angle $\theta$ varies smoothly from its far field preferred value to vanish on the $z$-axis. As $\theta$ vanishes, the helical integral curves degenerate to a straight line along the $z$-axis which, having no curvature, is a $\beta$ line. Using instead $\phi = qz + s \arctan(z/x)$ yields an edge dislocation, Fig.~\ref{fig:examples}(c). Here the $\beta$ line does not coincide  with the dislocation itself, where $\phi$ is singular ($y$-axis, marked D in Fig.~\ref{fig:examples}(c)). Instead it is displaced slightly to one side, at the position of the hyperbolic point (H in Fig.~\ref{fig:examples}(c)) where $\nabla\phi$ is zero~\cite{kamien2016}. A detailed comparision of the winding of $\phi$ and its singularities versus that of the bend vector is given in the Supplemental Material, which  contains Ref.~\cite{moffatt1992}. These examples can be set in a more general context that captures any smectic texture. For a smectic phase field $\phi$ with layer normal ${\bf N}$ we set ${\bf n} = \cos\theta \,{\bf N} + \sin\theta \bigl[ \cos\phi \,{\bf e}_1 + \sin\phi \,{\bf e}_2 \bigr]$, where ${\bf e}_1, {\bf e}_2$ are an orthonormal basis for the planes orthogonal to ${\bf N}$ that is parallel transported along it, $(\nabla_{{\bf N}} {\bf e}_1)\cdot{\bf e}_2=0$. The singularities in the smectic phase field then all induce $\beta$ lines in the director. In the Supplemental Material we provide examples for twist-grain-boundaries~\cite{Dozov2017,murachver2019,matsumoto2017} and parabolic focal conics~\cite{kleman2018,Alexander2010}. 

\begin{figure*}[t]
\centering
\includegraphics[width=\linewidth]{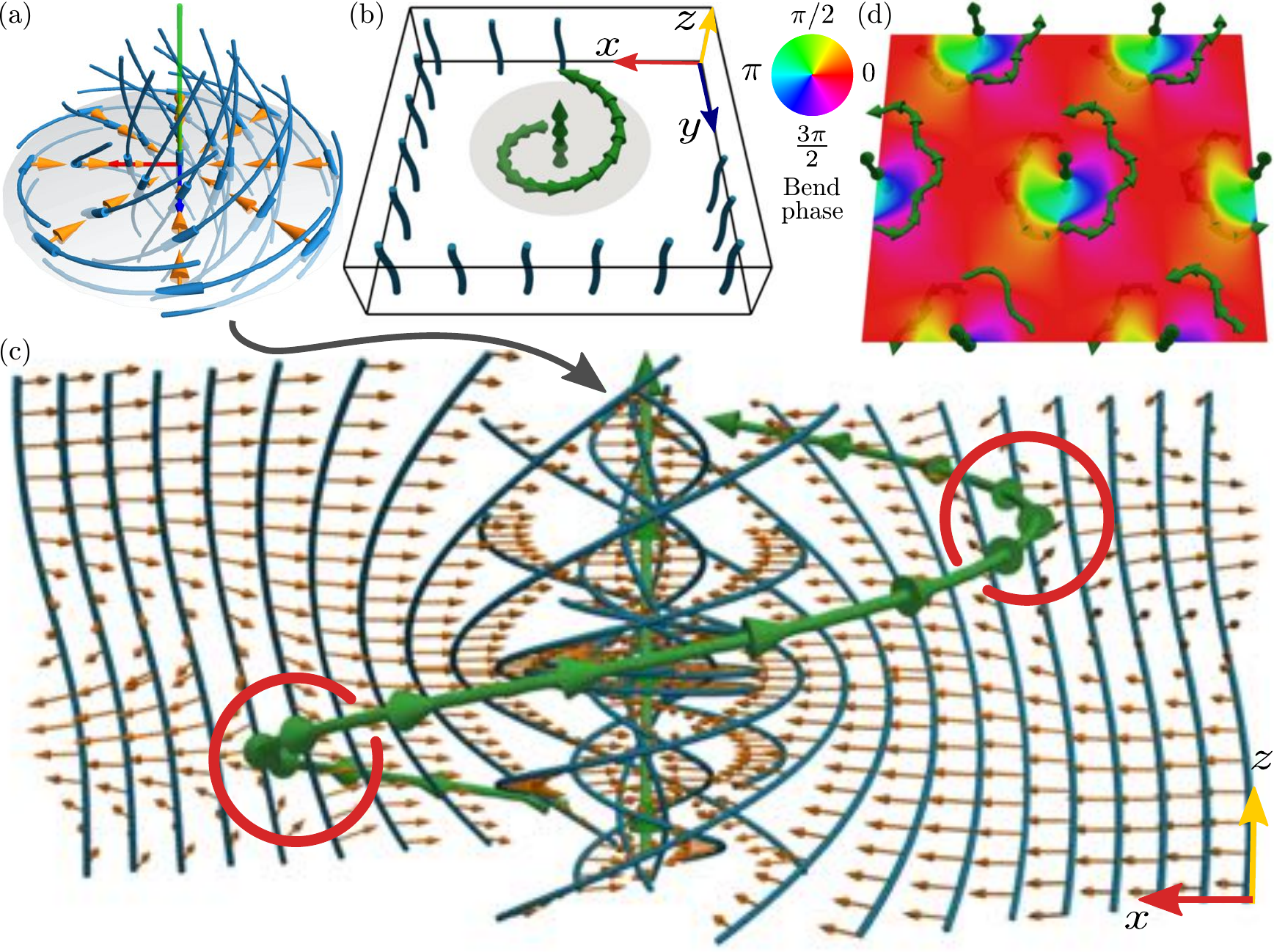}
    \caption{$\beta$ lines in double twist and Skyrmion vortex structures. (a) $\beta$ line at the centre of a double twist cylinder --- the integral curves of the director wind about the $\beta$ line, making this $+1$ defect topologically distinct from the screw dislocation of Fig.~\ref{fig:examples}(a). (b) A single Skyrmion, indicated by the grey disc, embedded in a heliconical background. There are two cooriented $\beta$ lines, which are topologically required by the Gauss-Bonnet-Chern theorem \eqref{eq:GBC}. Orientations are indicated by a choice of tangent vector along the $\beta$ lines, and agree with the far field heliconical director having negative $z$ component. (c) Detailed structure of the director integral curves (blue), bend vector (orange) and $\beta$ lines (green) of a twist-bend Skyrmion. Red circles highlight the winding of the bend vector about the second, helical, $\beta$ line. (d) A hexagonal lattice of twist-bend Skyrmions, with midplane coloured by $xy$ angle of the bend vector. Simulation results in panels (b,c,d) are shown for $\theta = 0.1, U/C =0.3$.}
\label{fig:skyrmion}
\end{figure*}

A separate set of examples of $\beta$ lines is provided by the vortex structures familiar from cholesterics, such as the axes of double twist cylinders or the cores of Skyrmions. A canonical example is the double twist profile ${\bf n} = \cos q\rho \,{\bf e}_z + \sin q\rho \,{\bf e}_{\phi}$ shown in Fig.~\ref{fig:skyrmion}(a), for which the bend is radial, ${\bf b} = - \frac{1}{\rho} \sin^2 q\rho \,{\bf e}_{\rho}$, and vanishes linearly along the axis with winding number $+1$. Although the winding number is the same as the $s=+1$ screw dislocation, Fig.~\ref{fig:examples}(a), the structure is distinct; each helical integral curve encircles the $\beta$ line, in contrast to the screw dislocation where they do not. This observation establishes that these two $\beta$ lines are topologically distinct, in the sense that one cannot convert one into the other without creating additional $\beta$ lines. Such a double twist cylinder occurs at the core of a (twist-bend nematic) Skyrmion embedded in a heliconical background, Fig.~\ref{fig:skyrmion}(b,c); Skyrmions are non-singular field configurations that are (meta)stable states in cholesterics and in ferromagnets with Dzyaloshinski-Moriya interaction~\cite{foster2019,ackerman2014,ackerman2017,duzgun2018,sutcliffe2017}, which carry a topological charge $Q=\frac{1}{4\pi} \int {\bf n} \cdot \partial_{x} {\bf n} \times \partial_{y} {\bf n} \,dx dy$. The analogous helical director structures immediately suggest it is possible Skyrmions also arise in twist-bend nematics, and indeed we find them to be (meta)stable in simulation. The Skyrmion contains two $\beta$ lines (Fig.~\ref{fig:skyrmion}(b,c)), one along the central axis with the structure of the double twist cylinder, Fig.~\ref{fig:skyrmion}(a), and the second a helix with pitch equal to that of the heliconical far field director and winding number of the bend equal to $-1$. These $\beta$ lines are a topological necessity and count the Skyrmion charge $Q$; we demonstrate below (and provide further detail in the Supplemental Material) that there are two $\beta$ lines per Skyrmion. Fig.~\ref{fig:skyrmion}(d) shows a hexagonal lattice of Skyrmions, again (meta)stable in simulation. A full Skyrmion phase diagram, analogous to that constructed for cholesterics~\cite{afghah2017}, would be of clear interest, although it is not the focus of this work; here, we simply note that we have confirmed (meta)stability for the heliconical far field angle $\theta \in [0.1,0.5]$, $U/C \in [0.1,0.5]$, in simulations perfomed using periodic boundary conditions with box height chosen to match one pitch length ($2\pi$ rotation) of the twist-bend director. The stability we have seen suggests that twist-bend Skyrmions could be directly nucleated by adapting techniques used in cholesteric cells or in magnetic systems.

\begin{figure*}[tb]
\centering
\includegraphics[width=0.99\linewidth]{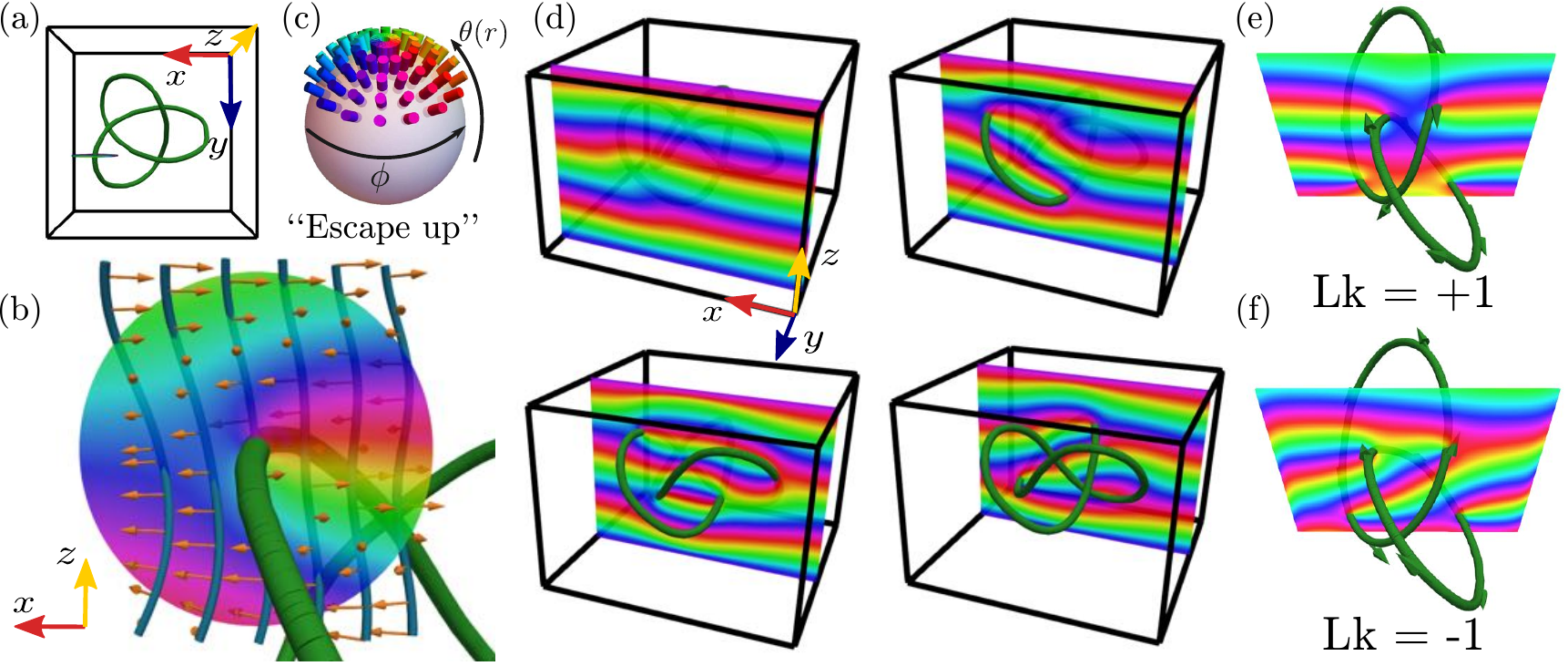}
    \caption{Knotted $\beta$ line meron textures in twist-bend nematics. (a) $\beta$ line tied into a trefoil knot, embedded in the heliconical background, with a cross-sectional slice marked and shown in (b). The local structure of the director (b,c) is an escape-up meron containing a single $\beta$ line (colour denotes bend phase; same as in Fig.~\ref{fig:skyrmion}). The director in the cross-section shown in (b) covers the top of the unit sphere as shown in (c), corresponding to escape up. In (d) we show the $\beta$ line from the side, with the bend phase (colour) on several cross-sectional slices. (e,f) Hopf links with linking numbers $+1$ (e) and $-1$ (f), and their distinct phase fields.}
    \label{fig:3D}
\end{figure*}

Thus far we have discussed $\beta$ lines in the context of particular examples motivated by experimentally relevant structures in the twist-bend nematic or in cholesterics. We now turn to a description of their geometric structure and topological significance in general --- we emphasise that our discussion is applicable to any vector or line field. In our simple examples, the director ${\bf n}$ is either colinear with the $\beta$ line tangent $\bf t$, as in Figs.~\ref{fig:examples}(a, b) and Fig.~\ref{fig:skyrmion}(a), or orthogonal to it as in Fig.~\ref{fig:examples}(c). However generically neither is the case, and ${\bf n}$ and ${\bf t}$ make some intermediate angle. Points where they are orthogonal have codimension one and are called Legendrian (see for example~\cite{geiges2008}); points of colinearity are codimension two and do not occur except in situations of high symmetry. A local description of a generic point on a $\beta$ line can be developed by introducing adapted coordinates ${\bf n} \approx n_x \,{\bf e}_x + n_y \,{\bf e}_y + {\bf e}_z$ and expanding in a Taylor series, retaining only terms that contribute at linear order to the bend:
\begin{align}
\begin{bmatrix} n_x \\ n_y \end{bmatrix} & = \biggl[ \Bigl. \nabla_\perp {\bf n} \Bigr\rvert_0 + z \Bigl. \bigl( \partial_z\nabla_\perp {\bf n} \bigr) \Bigr\rvert_0 \biggr] \begin{bmatrix} x \\ y \end{bmatrix} + \frac{1}{2} z^2 \begin{bmatrix} s_x \\ s_y \end{bmatrix} ,
\label{eq:director_local} \\
\begin{bmatrix}b_x \\ b_y\end{bmatrix} & = \biggl[ \Big( \Bigl. \nabla_\perp {\bf n} \Bigr\rvert_0 \Big)^2 + \Bigl. \partial_z\nabla_\perp {\bf n} \Bigr\rvert_0 \biggr] \begin{bmatrix} x \\ y \end{bmatrix} + z \begin{bmatrix} s_x \\ s_y \end{bmatrix} .
\label{eq:bendprofile}
\end{align}
Here $\nabla_{\perp} {\bf n} = \Bigl[ \begin{smallmatrix} \partial_x n_x & \partial_y n_x \\ \partial_x n_y & \partial_y n_y \end{smallmatrix} \Bigr]$ denotes the 2$\times$2 matrix of orthogonal gradients of the director~\cite{machon2016} and $\partial_z \nabla_{\perp} {\bf n}$ is its rate of change along the local director; $[s_x$, $s_y]$ controls the angle between ${\bf t}$ and $\bf n$. The winding number in the $xy$-plane is $\pm 1$ according to the sign of $\det\bigl( (\nabla_{\perp}{\bf n}|_{0})^2+\partial_z \nabla_{\perp}{\bf n}|_{0} \bigr)$. When the derivatives $\bigl. \partial_z \nabla_{\perp}{\bf n} \bigr|_{0}$ are negligible this reduces to $(\det \nabla_{\perp}{\bf n}|_{0})^2$ and the winding is always $+1$, so that the different profiles of $\beta$ lines are controlled crucially by the parallel derivatives of the orthogonal director gradients. In the Supplemental Material, which includes Refs.~\cite{berry1977,etnyre1999}, we use \eqref{eq:director_local}, \eqref{eq:bendprofile} to construct a variety of $\beta$ lines with different local profiles, including a Legendrian point.

We now describe the global structure beginning with a canonical orientation of $\beta$ lines via the operator $\nabla {\bf b}$. Along the $\beta$ lines there are two canonical frames; the tangent vector to the curve $\bf t$ and normal plane $\nu$, and the director $\bf n$ and its normal plane $\xi$. $\nabla {\bf b}$ defines an isomorphism $\nu\to\xi$, detailed in the Supplemental Material, and we orient the $\beta$ line such that this isomorphism preserves orientation. We note that the orientation obtained reverses upon ${\bf n}\to -{\bf n}$, which corresponds precisely to the change in sign of point defects (or Skyrmion charge) under the same replacement~\cite{alexander2012}. On the complement of the $\beta$ lines there is the Frenet-Serret framing of the director integral curves. The $\beta$ lines are singularities of this framing. We write ${\bf b} = \kappa \,{\bf e}_1$, with $\kappa$ the curvature of the integral curves, and ${\bf e}_2 = {\bf n} \times {\bf e}_1$. This framing yields a connection 1-form $\omega = (\nabla {\bf e}_1) \cdot {\bf e}_2$ for the plane field $\xi$. The component of $\omega$ along the director is the torsion $\tau = \omega({\bf n}) = (\nabla_{{\bf n}} {\bf e}_1) \cdot {\bf e}_2$, while the vector dual to it is the local pitch axis of the heliconical twist-bend state. For example, the smectic-based twist-bend director~\eqref{eq:tb_smectic} has connection 1-form $\omega = \cos\theta \,d\phi$; the torsion is $\tau \approx q \cos^2\theta$ and the pitch axis is along $\nabla\phi$. Topological information is conveyed by the associated curvature 2-form $\Omega = - \sin\theta \,d\theta \wedge d\phi = \frac{-1}{2} \epsilon_{ijk} n_{i} dn_j \wedge dn_k$.
Given a surface $S$, the $\beta$ lines intersect it in a set of points $p_i$ and by the Gauss-Bonnet-Chern theorem
\begin{equation}
\frac{1}{2\pi} \int_{\partial S} \omega - \frac{1}{2\pi} \int_{S} \Omega = \sum_{i} \mathrm{Int}_{p_i}(\beta,S) ,
\label{eq:GBC}
\end{equation}
where $\mathrm{Int}_{p_i}(\beta,S)$ denotes the signed intersection number at point $p_i$ of an oriented $\beta$ line with an oriented surface $S$. For Skyrmion textures this total intersection number is $2Q$, giving two $\beta$ lines per Skyrmion as seen in Fig.~\ref{fig:skyrmion}. In the Supplemental Material, which contains also Refs.~\cite{ackerman2017prx,chen2013prl,sutcliffe2018,calugareanu1961,fuller1971}, we give detailed applications of \eqref{eq:GBC} to the examples of the screw dislocation (Fig.~\ref{fig:examples}), the double twist cylinder (Fig.~\ref{fig:skyrmion}(a)), and the Skyrmion texture (Fig.~\ref{fig:skyrmion}(b,c)), in each case demonstrating the topological necessity of the $\beta$ lines present.

Finally, we discuss fully three-dimensional textures and show that it is possible to embed an arbitrary knotted or linked set of $\beta$ lines into a heliconical background, via an extension of our constructions for screw and edge dislocations. Given any knot or link $K$, the director
\begin{equation}
{\bf n} = \cos\theta \,{\bf e}_z + \sin\theta \bigl[ \cos\phi_K \,{\bf e}_x + \sin\phi_K \,{\bf e}_y \bigr] ,
\label{eq:meron}
\end{equation}
where $\phi_{K} = qz + \frac{1}{2} \omega_{K}$, with $\omega_K$ the solid angle function for $K$~\cite{binysh2018}, embeds a helical winding of the director integral curves around a tubular neighbourhood of $K$, Fig.~\ref{fig:3D}; as before, the cone angle $\theta$ should be made to vary from its far field value to vanish along $K$. The phase winding in the helical integral curves guarantees the existence of a $\beta$ line. Examples for the trefoil knot, Fig.~\ref{fig:3D}(a-d), and Hopf link, Fig.~\ref{fig:3D}(e,f), are shown. 

The director texture around the knot is that of a meron tube extruded along $K$. A meron is a fractionalisation of a Skyrmion that carries half the topological charge~\cite{duzgun2018,yu2018}. $\beta$ lines provide a natural geometric perspective on this fractionalisation: since each Skyrmion comprises two $\beta$ lines, a single $\beta$ line represents half a Skyrmion, {\sl i.e.} a meron.
In terms of the heliconical phase field, $\phi_{K}$, these meron tubes are edge dislocations where heliconical layers terminate. Exactly these structures were recently created experimentally in cholesteric cells and shown to form highly controllable and responsive knotted solitons~\cite{tai2019}. In that experiment, links of `escape up' and `escape down' meron tubes combined to give non-zero Hopf invariant. For the twist-bend nematic phase, the small conical angle ($\theta \approx 25^{\circ}$~\cite{chen2013}) gives an energetic preference to `escape up' merons over `escape down', whereas in cholesterics ($\theta = \pi/2$) the two types of meron are degenerate. Even with only `escape up' merons, where the Hopf invariant is trivial, the linking is still a relevant quantity with distinct textures for different values of the linking number, $\mathrm{Lk}(\beta_i,\beta_j)$. In Fig.~\ref{fig:3D}(e,f) we show examples for the Hopf link with linking numbers $\pm 1$ where the layer structure through the middle of the link is different in the two cases; further images of an unknot, the trefoil, and both types of Hopf link are given in the Supplemental Material.

The triviality, or otherwise, of the Hopf invariant can also be seen just from the $\beta$ lines and the invariant $\Theta = \sum_{i} s_i^2\mathrm{SL} ( \beta_i ) + \sum_{i\neq j} s_i s_j\mathrm{Lk} ( \beta_i , \beta_j )$, familiar from helicity and abelian Chern-Simons theory~\cite{ArnoldKhesin}. This is not directly equal to the Hopf invariant, but is an alternative way of presenting the homotopy group $\pi_3(S^2) \cong Z$ that describes three-dimensional solitons~\cite{Gompf}. The integer $s_j$ denotes the strength of the $j$th $\beta$ line $\beta_j$, generically $\pm 1$, and the self-linking number, $\mathrm{SL}(\beta)$, is defined as follows: consider the total rotation $\int_{B^{\prime}} {\bf e}_2 \cdot d{\bf e}_1$ of the Frenet-Serret frame about the director along any push-off $B^{\prime}$ giving a zero-framing for the $\beta$ line. Part of this rotation is an intrinsic Berry phase $\gamma$, equal to the area on the unit sphere bound by the curve traced out by {\bf n} along $B$. The difference $\gamma - \int_{B^{\prime}} {\bf e}_2 \cdot d{\bf e}_1 = 2\pi \, \textrm{SL}(\beta)$ defines the self-linking. See the Supplemental Material for an illustration of how to compute the self-linking number.

We have given an initial description of geometric degeneracies in the bend of a vector field, which we call $\beta$ lines, and their connection to topological features, including smectic singularities, Skyrmions and merons. We have couched the majority of the discussion around the twist-bend nematic phase, in which the $\beta$ lines are novel topological defects, however the same structures arise in any orientationally ordered material. As one example in a different setting, active materials with extensile activity exhibit a bend driven instability in (three-dimensional) active nematics and cholesterics~\cite{whitfield2017,binysh2020,duclos2019} and so naturally exist in states with non-zero bend distortion. The geometric degeneracies we have introduced here will also arise there and may provide a means for their analysis.

\tocless\acknowledgements{This work was supported by the UK EPSRC through Grant No. EP/L015374/1. JB supported by a Warwick IAS Early Career Fellowship.}

\pagebreak
\onecolumngrid
\beginsupplement

\begin{center}
{\large {\bf Supplemental Material}}
\end{center}

\tableofcontents

\section*{Introduction}

This supplemental material provides, firstly, a detailed account of the bend distortions of a unit vector field, their geometric degeneracies and the topological information they carry. The results obtained apply to any material or physical system described (even in part) by such a unit vector, or line, field. In the same way that the geometry of twist is used primarily in the context of chiral phases, such as cholesteric liquid crystals, helimagnets and Beltrami flows, where the natural state is one of non-zero twist, we anticipate that the geometry of bend will be used primarily for phases where the natural state is one of non-zero bend. For this reason, we illustrate our general discussion with examples taken from a concrete, minimal example of such a system, the twist-bend nematic --- in the same way that cholesterics illustrate general features of the geometry of twist, we shall use the twist-bend nematic to illustrate general aspects of the geometry of bend. As such, the specific free energy we shall use to model the twist-bend nematic, drawn from recent literature \cite{shamid2013,jakli2018}, will be of subsidiary importance. The reader interested in the general structure of bend distortions may consult \S\S\ref{sec:general},\ref{sec:Local},\ref{sec:FrenetSerret},\ref{sec:Meron} and read these independently of any consideration of the twist-bend nematic, whose modelling and free energy is discussed briefly in \S\ref{sec:FreeEnergy}.

The second purpose of this supplement is to provide additional graphical renderings and mathematical detail of the various defects in twist-bend nematics discussed in the main text --- of the screw and edge dislocations \S\S\ref{sec:Heliconical},\ref{subsec:screw},\ref{subsec:edge}, of Skyrmions and Skyrmion lattices \S\ref{sec:Skyrmion}, and of three-dimensional knotted merons \S\ref{sec:Meron}. In addition, we discuss two smectic-like defects not detailed in the main text, twist grain boundaries \S\ref{sec:TGB} and focal conics \S\ref{sec:FocalConic}.

\subsection{Geometry of Orientational Order}
\label{sec:general}

\begin{figure}[tb]
\centering
\includegraphics[width=0.8\linewidth]{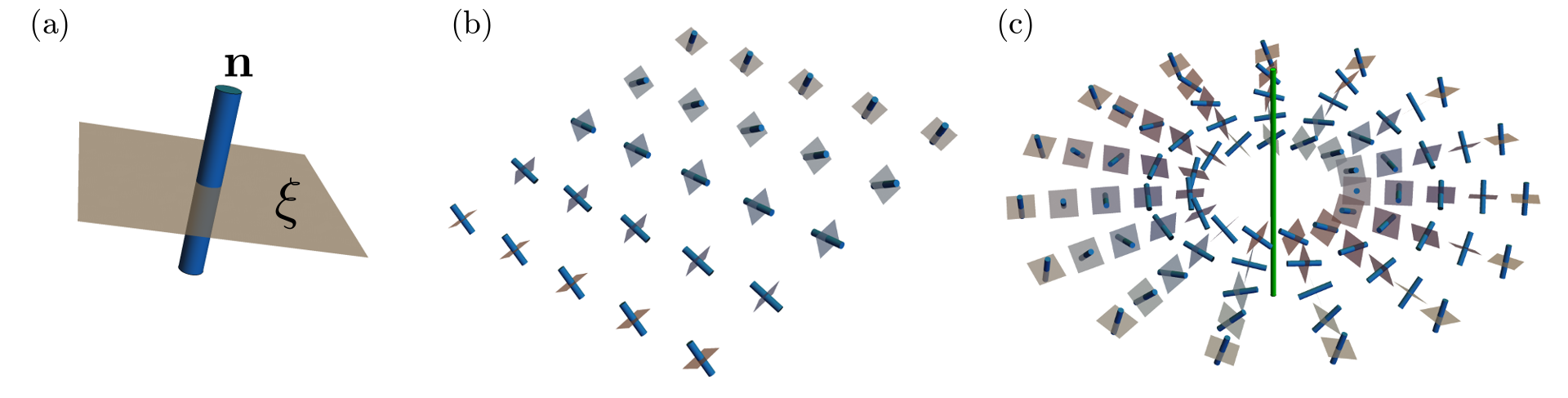}
    \caption{An illustration of the plane field $\xi$ associated to a director ${\bf n}$. (a) At each point, $\xi$ is defined to be the orthogonal plane (grey) to the director (blue). ${\bf n}$ and $\xi$ are shown for (b) the cholesteric ground state and (c) a double-twist cylinder, whose axis is indicated by the green line.}
\label{fig:planes}
\end{figure}

The geometrical description of orientational order comes from a natural decomposition of the director gradients~\cite{machon2016,alexander2018,machon2019}. The director gives a canonical splitting of directions in space at each point, into those parallel to the director and those perpendicular to it. The latter define a two-dimensional vector space at every point, called the orthogonal plane field, $\xi$. Fig.~\ref{fig:planes} illustrates this splitting and the plane field $\xi$ at a single point in space (Fig.~\ref{fig:planes}(a)), for the cholesteric ground state (Fig.~\ref{fig:planes}(b)), and for a double-twist cylinder (Fig.~\ref{fig:planes}(c)). The local symmetry of the director field gives an action of a subgroup of the rotation group isomorphic to $SO(2)$ under which the director gradients naturally split as
\begin{equation}
\partial_i n_j = n_i (n_k \partial_k) n_j + \frac{\nabla\cdot{\bf n}}{2} \bigl( \delta_{ij} - n_i n_j \bigr) + \frac{{\bf n}\cdot\nabla\times{\bf n}}{2} \,\epsilon_{ijk} n_k + \Delta_{ij} .
\label{Seq:director_gradients}
\end{equation}
The first term gives the derivatives parallel to the director field, $\nabla_{\parallel}{\bf n}$, and the remainder the orthogonal gradients, $\nabla_{\perp}{\bf n}$. The orthogonal gradients can be thought of as a linear transformation on the orthogonal plane field --- the shape operator for the director field --- defined by ${\bf v} \mapsto ({\bf v}\cdot\nabla){\bf n}$ for any orthogonal vector ${\bf v}$. The first two terms in the orthogonal gradients are isotropic and contain the splay and twist distortions, while the last term, $\Delta_{ij}$, is the anisotropic part of the orthogonal gradients; it is a traceless, symmetric, linear transformation on $\xi$ that transforms as a spin $2$ object under the action of the local $SO(2)$ symmetry group. Its eigenvectors are the directions of principal curvature of the director field. Closely related to it is the linear transformation $\Pi_{ij} = \Delta_{il} \epsilon_{ljk} n_k$, the anisotropic part of the chirality pseudotensor, whose principal eigenvector defines the pitch axis in cholesterics and helimagnets~\cite{machon2016,alexander2018}. Of course, $\Delta$ and $\Pi$ are defined for any type of orientational order and not only for cholesterics, but in cholesterics where non-zero twist is energetically preferred they gain added significance. The defects in the pitch axis --- called $\lambda$ lines and readily visible under optical microscopy --- correspond to the zeros of $\Delta$ (and equivalently $\Pi$)~\cite{machon2016,alexander2018}. In the general case, the zeros of $\Delta$ (equivalently $\Pi$) are the umbilics of the director field, where the orthogonal gradients are locally isotropic. As these are zeros of a linear transformation on a vector space they carry topological information~\cite{machon2016,alexander2018,machon2019}; specifically, they identify the topology of the director field modulo elements of order $4$~\cite{machon2016,machon2016prsa,machon2019}.

The parallel gradients, $\nabla_{\parallel}{\bf n}$, define the bend distortion ${\bf b} := ({\bf n}\cdot\nabla){\bf n} = - {\bf n} \times (\nabla\times{\bf n})$; its geometric interpretation is that it is the curvature of the integral curves of the director field.  The bend is a vector that is everywhere orthogonal to the director field, ${\bf b}\cdot{\bf n} = 0$, and is therefore a section of the orthogonal plane field $\xi$. As a section of a rank $2$ vector bundle, $\bf b$ has zeros of codimension $2$ which form one-dimensional curves within the texture. We call these curves $\beta$ lines; they are the central object of the present work. These $\beta$ lines are the locus of inflection points in the director integral curves; as a director integral curve intersects a $\beta$ line, the curvature of the director integral curve vanishes. $\beta$ lines furnish a geometric fingerprint of the director field, reflecting its geometric structure while also conveying topological information by representing the Poincar\'e dual to the Euler class of $\xi$; we shall see numerous examples of this in the following sections.

\subsection{Free Energy}
\label{sec:FreeEnergy}

The description of director gradients given so far has been fully general, applying to any form of orientational order and, in fact, independent of any energetic considerations. However, energetic considerations are also important as they will constrain the facets of the geometry that are most important in determining the properties of different phases. The decomposition of director gradients~\eqref{Seq:director_gradients} leads immediately to the Frank free energy~\cite{Machon,machon2019}
\begin{equation}
F = \int \biggl\{ \frac{\tilde{K}_{1}}{2} \bigl( \nabla \cdot {\bf n} \bigr)^2 + \frac{\tilde{K}_{2}}{2} \bigl( {\bf n}\cdot\nabla\times{\bf n} \bigr)^2 + \frac{\tilde{K}_{3}}{2} \bigl| ({\bf n}\cdot\nabla) {\bf n} \bigr|^2 + \tilde{K}_{4} \bigl| \Delta \bigr|^2 \biggr\} dV ,
\label{Seq:Frank}
\end{equation}
where $|\Delta|^2 = \Delta_{ij} \Delta_{ij}$ and the $\tilde{K}_i$ are elastic moduli in terms of which the usual Frank constants are~\cite{Machon,machon2019,selinger2019}
\begin{align}
& K_1 = \tilde{K}_1 + \tilde{K}_4 , && K_2 = \tilde{K}_2 + \tilde{K}_2 , && K_3 = \tilde{K}_3 , && K_{24} = \tilde{K}_4 .
\end{align}
A lucid exposition of this approach to the Frank free energy along with insightful applications to interpreting liquid crystal textures is given in~\cite{selinger2019}.

Beyond the nematic phase, different types of orientational order emphasise particular aspects of the geometry by energetically favouring a non-zero value for one of the four parts of the director gradients~\eqref{Seq:director_gradients}. The most familiar case is that of cholesterics where the twist term in the free energy~\eqref{Seq:Frank} becomes $\frac{1}{2} \tilde{K}_2 ({\bf n}\cdot\nabla\times{\bf n} + q_0)^2$, with $q_0$ the chirality.
The other three parts of the decomposition~\eqref{Seq:director_gradients} do not provide scalar invariants of the director field (invariant under the nematic symmetry ${\bf n} \sim -{\bf n}$). Energetic terms promoting a non-zero value for these geometric distortions can be given in the Brazovskii form
\begin{align}
& \frac{\tilde{K}_1}{2} \Bigl( \bigl| \nabla \cdot {\bf n} \bigr|^2 - s_0^2 \Bigr)^2 , && \frac{\tilde{K}_3}{2} \Bigl( \bigl| ({\bf n} \cdot \nabla) {\bf n} \bigr|^2 - b_0^2 \Bigr)^2 , && \tilde{K}_4 \Bigl( \bigl| \Delta \bigr|^2 - \Delta_0^2 \Bigr)^2 ,
\end{align}
where $s_0$ is the preferred magnitude of the splay, $b_0$ is the preferred magnitude of the bend and $\Delta_0$ is the preferred magnitude of the anisotropic part of the orthogonal gradients, although these are far from the most general expressions. An alternative description creates non-zero values for the splay, bend or anisotropic orthogonal gradients by introducing auxiliary fields and couplings of the form
\begin{align}
& -\lambda \bigl( n_i \partial_j n_j \bigr) p_i , && -\lambda \bigl( n_j \partial_j n_i \bigr) p_i , && -\lambda \Delta_{ij} T_{ij} ,
\end{align}
respectively. This is the approach originally suggested by Meyer for describing spontaneously modulated splay and bend phases and adopted by the Kent State group~\cite{shamid2013}. The relationship between the two approaches has been described in the recent review~\cite{jakli2018}. In the case of the twist-bend nematic the free energy can be taken to have the form
\begin{equation}
F = \int \biggl\{ \frac{\tilde{K}_1}{2} \bigl( \nabla \cdot {\bf n} \bigr)^2 + \frac{\tilde{K}_2}{2} \bigl( {\bf n} \cdot \nabla \times {\bf n} \bigr)^2 + \frac{\tilde{K}_3}{2} \bigl| ({\bf n}\cdot\nabla) {\bf n} \bigr|^2 + \tilde{K}_4 \bigl| \Delta \bigr|^2 - \lambda \bigl[ ({\bf n}\cdot\nabla) {\bf n} \bigr] \cdot {\bf p} + \frac{C}{2} \bigl| \nabla {\bf p} \bigr|^2 + \frac{U}{4} \bigl( 1 - |{\bf p}|^2 \bigr)^2 \biggr\} dV ,
\label{Seq:TB_energy}
\end{equation}
where $C$ is an elastic modulus for the auxiliary polarisation field ${\bf p}$ and $U$ sets the scale of its bulk ordering energy. In ordinary nematics (and cholesterics) the Frank free energy is often simplified by adopting a one-elastic-constant approximation, replacing the four Frank elastic terms with the single term $\frac{K}{2} |\nabla{\bf n}|^2$, and we adopt this reduced form for simplicity in generating numerical examples.

Our focus here is on geometric and topological properties of the director field, which are largely insensitive to the exact form of the free energy and have general applicability for typical values of material constants.

\section*{Basic Examples of Bend Geometry and $\beta$ Lines}

In the following sections we detail the construction of the basic defects in twist-bend nematics discussed in the main text, as well as give additional graphical renderings of these textures from different perspectives and with different features emphasised. The heliconical ground state of the twist-bend nematic has one-dimensional periodic spatial modulation. On scales large compared to the heliconical pitch, its elastic deformations and hydrodynamic modes are the same as those of a smectic~\cite{kamien1996,parsouzi2016,meyer2016}, as is the case also for cholesterics~\cite{radzihovsky2011}. The polarisation is a non-hydrodynamic mode~\cite{parsouzi2016}. As such, many calculations from the literature on smectics can be applied directly to give a coarse description of the energetics of defects and textures in twist-bend nematics, when the latter are closely similar to known smectic textures. Our focus will be on describing these states from the twist-bend perspective where they may be visualised as disruptions to the family of helices which make up the director integral curves. We begin with a recapitulation of the geometry of the heliconical ground state, \S\ref{sec:Heliconical}, before discussing the two basic smectic-like defects introduced in the main text --- screw (\S\ref{subsec:screw}) and edge (\S\ref{subsec:edge}) dislocations. We also detail two more complex smectic-like defects not presented in the main text, grain boundary phases (\S\ref{sec:TGB}) and focal conics (\S\ref{sec:FocalConic}). We then described examples related to isolated Skyrmions and Skyrmion lattices (\S\ref{sec:Skyrmion}), providing enlarged renderings of these textures to convey their complex structure.

\subsection{Heliconical State}
\label{sec:Heliconical}

\begin{figure}[tb]
\centering
\includegraphics[width=0.99\linewidth]{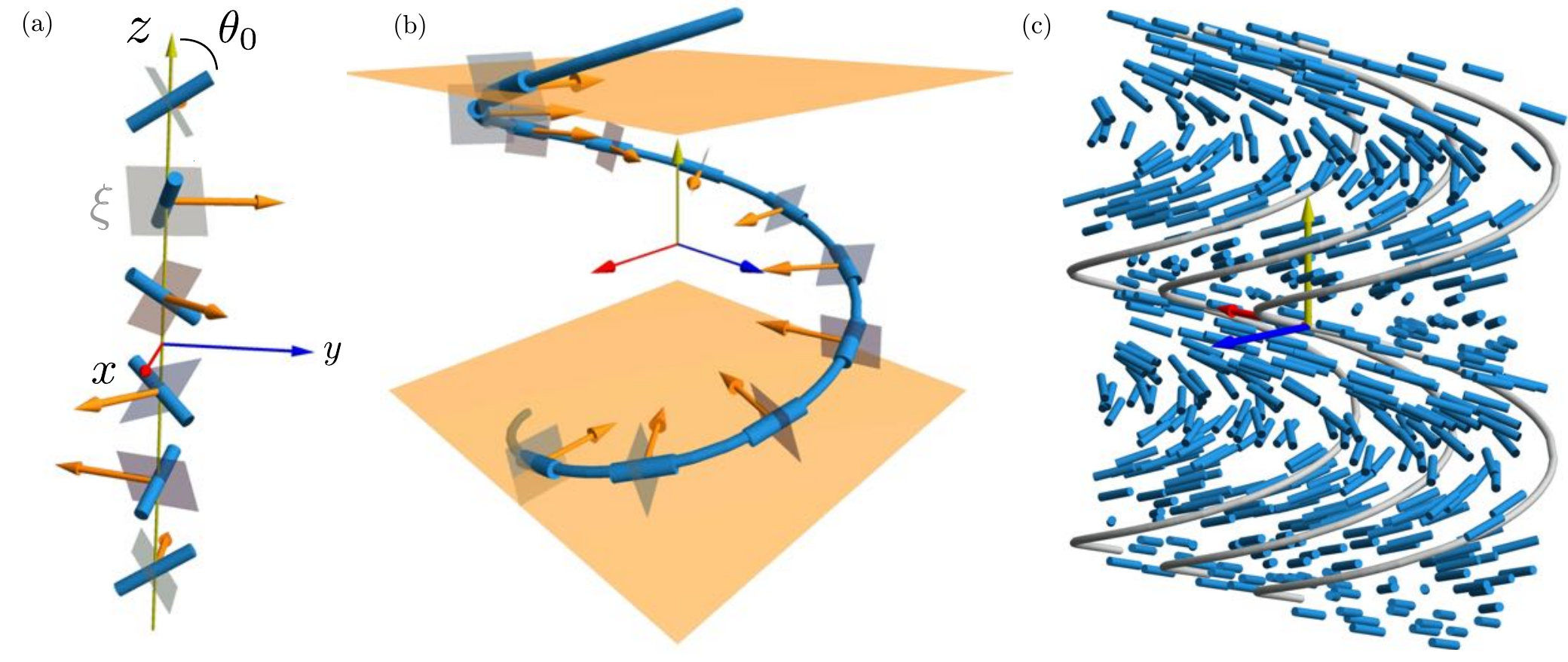}
    \caption{The heliconical texture. (a) The director ${\bf n}$ makes a constant angle $\theta_0 \in [0,\pi/2]$ with the $z$ axis and rotates with wavevector $q$ in the $xy$ plane. The bend $\bf b$ lies in the $xy$ plane, again rotating with wavevector $q$. (b) The integral curves of the director are helices; orange surfaces indicate a full turn of the helix, with pitch $2\pi/q$. (c) The director field fills space, giving a family of interlocking integral helices --- three such helices are shown in grey.}
\label{fig:Heliconical}
\end{figure}

The heliconical state can be given the following purely geometrical description. It is characterised by having a bend distortion of constant non-zero magnitude. The bend is the curvature of the director integral curves; curves with constant magnitude of curvature are helices. Taking the helical axis to be $z$, a general helical integral curve can be written
\begin{equation}
{\bf X}(z) = x_0 \,{\bf e}_x + y_0 \,{\bf e}_y + z \,{\bf e}_z + \frac{\tan\theta_0}{q} \Bigl[ \sin qz \,{\bf e}_x + (1 - \cos qz) \,{\bf e}_y \Bigr] ,
\label{Seq:helices}
\end{equation}
where $x_0, y_0$ are constants corresponding to the point in the $xy$-plane that the helix passes through. The helix has curvature $q \sin\theta_0 \cos\theta_0$ and torsion $q \cos^2\theta_0$; the unit tangent gives the director field of a heliconical state
\begin{equation}
{\bf n} = \cos\theta_0 \,{\bf e}_z + \sin\theta_0 \bigl[ \cos qz \,{\bf e}_x + \sin qz \,{\bf e}_y \bigr] ,
\label{Seq:heliconical}
\end{equation}
where $q$ is the helical wavevector and $\theta_0$ is the constant cone angle the director makes with the heliconical pitch axis --- here the $z$-axis. As $\theta_0 \to 0$ the director limits to the uniform orientation ${\bf e}_z$, with straight integral curves. When $\theta_0 \to \pi/2$ we recover the cholesteric ground state, and again the integral curves are straight lines, which now rotate uniformly as one moves along $z$.  In Fig.~\ref{fig:Heliconical} we show the heliconical texture \eqref{Seq:heliconical} and its helical integral curves \eqref{Seq:helices} for a generic cone angle, intermediate between these two extremes.

To analyse the director gradients we introduce the basis
\begin{align}
{\bf s}_1 & = - \sin qz \,{\bf e}_x + \cos qz \,{\bf e}_y , \\
{\bf s}_2 & = \sin\theta_0 \,{\bf e}_z - \sin\theta_0 \bigl[ \cos qz \,{\bf e}_x + \sin qz \,{\bf e}_y \bigr] ,
\end{align}
of the orthogonal planes $\xi$; these correspond to the normal and binormal vectors in the Frenet-Serret frame of the helical integral curves~\eqref{Seq:helices}. The director gradients are
\begin{equation}
\nabla {\bf n} = q \sin\theta_0 \cos\theta_0 \,{\bf n}\otimes{\bf s}_1 - \frac{q\sin^2\theta_0}{2} \bigl[ {\bf s}_1 \otimes {\bf s}_2 - {\bf s}_2 \otimes {\bf s}_1 \bigr] + \frac{q\sin^2\theta_0}{2} \bigl[ {\bf s}_1 \otimes {\bf s}_2 + {\bf s}_2 \otimes {\bf s}_1 \bigr] ,
\end{equation}
and we can read off that the bend is ${\bf b} = ({\bf n}\cdot\nabla){\bf n} = q\sin\theta_0 \cos\theta_0 \,{\bf s}_1$, the splay is $\nabla\cdot{\bf n} = 0$ and the twist is ${\bf n}\cdot\nabla\times{\bf n} = -q\sin^2\theta_0$. The anisotropic orthogonal gradients are
\begin{align}
& \Delta = \frac{q\sin^2\theta_0}{2} \bigl[ {\bf s}_1 \otimes {\bf s}_2 + {\bf s}_2 \otimes {\bf s}_1 \bigr] , && \Pi = - \frac{q\sin^2\theta_0}{2} \bigl[ {\bf s}_1 \otimes {\bf s}_1 - {\bf s}_2 \otimes {\bf s}_2 \bigr] ,
\end{align}
and from the linear transformation $\Pi$ we can read off that the cholesteric pitch axis is ${\bf s}_2$. We note that this is not the same as the heliconical pitch axis ($z$-axis); we explain how to identify the latter in \S\ref{sec:FrenetSerret}.

This description has emphasised the geometry of the heliconical state, independent of specific energetic considerations. Several free energies have been developed that have the heliconical director~\eqref{Seq:heliconical} as a ground state; some examples include~\cite{kamien1996,dozov2001,shamid2013,kats2014,pajak2018}. For the free energy~\eqref{Seq:TB_energy}, taking the limit $U\to\infty$ (which enforces $|{\bf p}|=1$), the preferred values of the heliconical cone angle $\theta_0$ and wavevector $q$ are
\begin{align}
& \cos 2 \theta_0 = 1+ \frac{2C}{K} - \biggl[ \frac{4C}{K} \biggl( 1+ \frac{C}{K} \biggr) \biggr]^\frac{1}{2} , && q = \frac{2 \lambda}{K} \cot 2 \theta_0 .
\end{align}

\subsection{Screw Dislocations}
\label{subsec:screw}

\begin{figure}[tbp]
\centering
\includegraphics[width=0.8\linewidth]{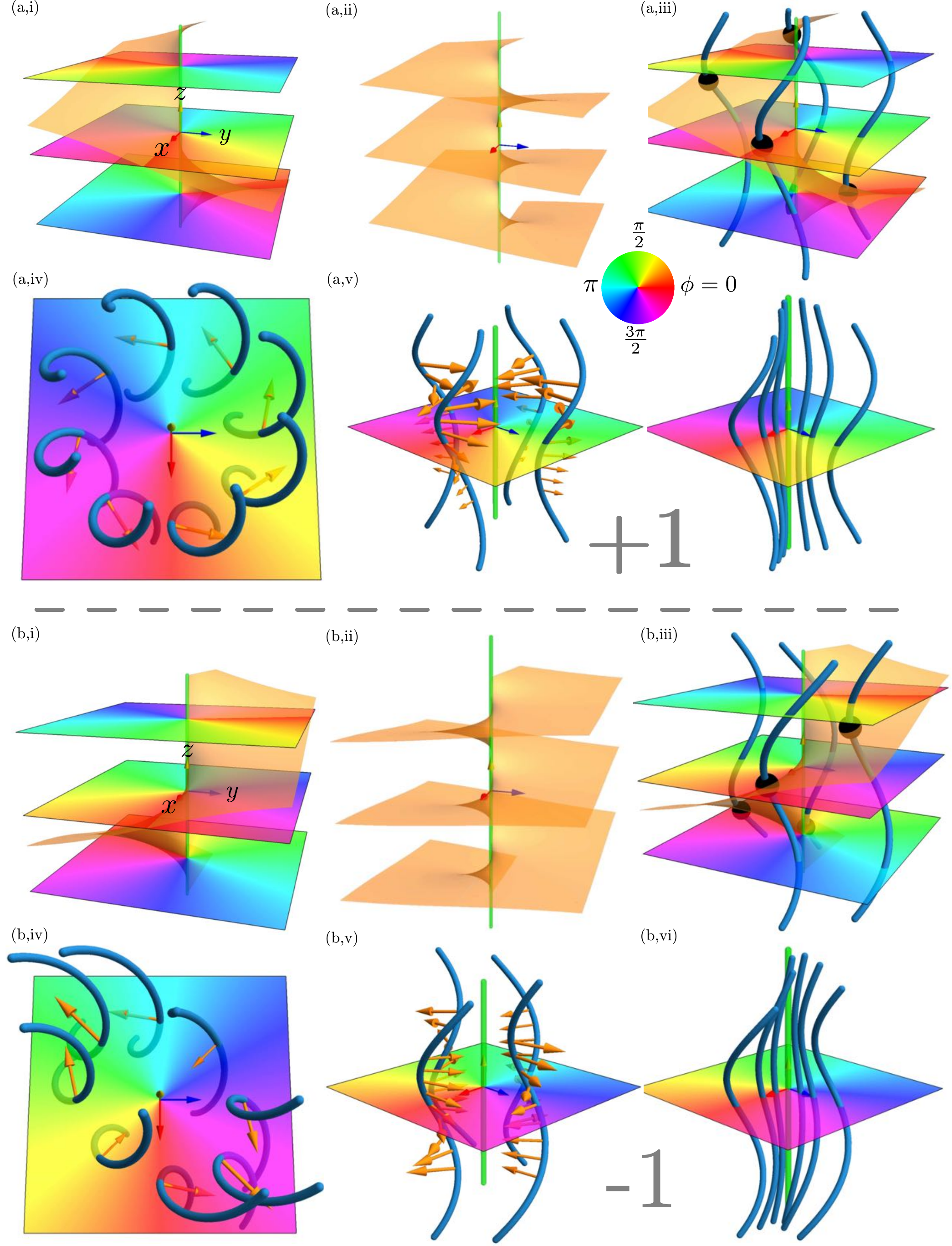}
     \caption{Screw dislocations in twist-bend nematics. Panels (a,b) show $+1,-1$ strength screw dislocations respectively. The $\beta$ line along the $z$ axis is shown in green. (a,b)(i) Helical phase field on three $z$ slices, with $\phi=0$ level set shown in orange. (a,b)(ii) Zoomed out view of $\phi = 0$ level set, showing equispaced layers away from the screw dislocation. (a,b)(iii) Director integral curves (blue) with their intersection with $\phi=0$ shown as black dots. (a,b)(iv) Top down view of integral curves, with bend vector (orange) shown on a $z=0$ slice. (a,b)(v) Perspective view of integral curves and their bend vector, showing periodic variation of the bend along the screw. (a,b)(vi) Degeneration of the integral curves to a straight line, which is also the $\beta$ line, as we approach the $z$ axis.}
\label{fig:Screw}
\end{figure}

The one-dimensional periodicity of the heliconical phase leads to a general correspondence with the elasticity of smectics and so a description in terms of `smectic-like' phase fields. The heliconical phase $\phi = qz$ in \eqref{Seq:heliconical} is the same as the phase in the mass-density wave of the smectic ground state. Other smectic phase fields --- corresponding to screw dislocations, edge dislocations, TGB phases, focal conics etc. --- lend themselves to analogous twist-bend states with the same helical phase field and provide examples of smectic-like defects in twist-bend nematics. We emphasise at the outset, however, that this is merely one class of defect in twist-bend nematics; the Skyrmion-type textures we describe later are not derived in this way from a smectic counterpart.

Our first example of a smectic-like defect is the screw dislocation, for which we consider the texture
\begin{equation}
    {\bf n} = \cos\theta(\rho) \,{\bf e}_z + \sin\theta(\rho) \bigl[ \cos\phi \,{\bf e}_x + \sin\phi \,{\bf e}_y \bigr] ,
\label{eq:Screw}
\end{equation}
where $\phi = qz + s\arctan(y/x)$, with $s = \pm 1, \pm 2, \dots $ the defect strength, and $\theta(\rho)$ interpolates smoothly from $0$ at the origin to the heliconical far field angle as $\rho := \sqrt{x^2+y^2} \to \infty$. In Fig.~\ref{fig:Screw} we show these textures for $s= +1,-1$ in panels (a,b) respectively. The phase field $\phi$ contains a smectic screw disclocation along the $z$ axis such that around any positively oriented loop in the $xy$-plane encircling the axis $\phi$ winds by $2\pi s$. This is shown by the winding colour map in Figs.~\ref{fig:Screw}(a,b)(i), which also show the level set $\phi = 0$ as an orange surface; this surface corresponds to the layers of a smectic screw dislocation. Note the difference in the sense of rotation between panels (a) and (b). Figs.~\ref{fig:Screw}(a,b)(ii) show the same level set $\phi =0$ but zoomed out, emphasising that away from the screw dislocation we simply have equally spaced layers, $\phi \approx qz$. In Fig.~\ref{fig:Screw}(a,b)(iii) we add integral curves of the director, with their intersection with the $\phi=0$ surface indicated by black points; in the limit $\rho \rightarrow \infty$ the integral curves are exactly helices and the marked points are locations along the integral curves of the same `helical phase'. The screw dislocation corresponds to a $2\pi s$ `phase slip', as can be seen in Fig.~\ref{fig:Screw}(a,b)(iv) in which we show a top down view of the integral curves alongside the phase $\phi$ on the $xy$ plane.

The bend of \eqref{eq:Screw} is
\begin{equation}
  {\bf b} = ({\bf n} \cdot \nabla \theta)\left[ \cos\theta \left( \cos \phi \,{\bf e}_x + \sin \phi \,{\bf e}_y \right) - \sin\theta \,{\bf e}_z \right] + ({\bf n} \cdot \nabla \phi)\sin\theta\left[-\sin\phi \,{\bf e}_x + \cos \phi \,{\bf e}_y \right].
  \label{eq:Bend}
\end{equation}
We first consider its far field behaviour. As $\rho \rightarrow \infty$, $\nabla \theta \rightarrow 0$ and \eqref{eq:Bend} becomes
\begin{equation}
  {\bf b} = q \cos\theta_0 \sin\theta_0 \left[ -\sin\phi \,{\bf e}_x + \cos \phi \,{\bf e}_y \right],
  \label{eq:BendFarField}
\end{equation}
exactly the heliconical bend but with $qz \rightarrow \phi = qz + \arctan(y/x)$. We conclude that the bend winds as $\phi$, and so there is a $2\pi s$ winding of the bend vector about the origin. This winding is shown in Fig.~\ref{fig:Screw}(a,b)(iv). The bend~\eqref{eq:BendFarField} also rotates along the pitch axis $z$ with pitch $2\pi/q$, giving a periodic structure to these defects along $z$, as shown in Figs.~\ref{fig:Screw}(a,b)(v). For $s=+1$, a radial profile rotates to become azimuthal and then back to radial. For $s=-1$, the axes of the $-1$ profile rotate along $z$.

\begin{figure}[t]
  \centering
  \includegraphics[width=1.0\linewidth]{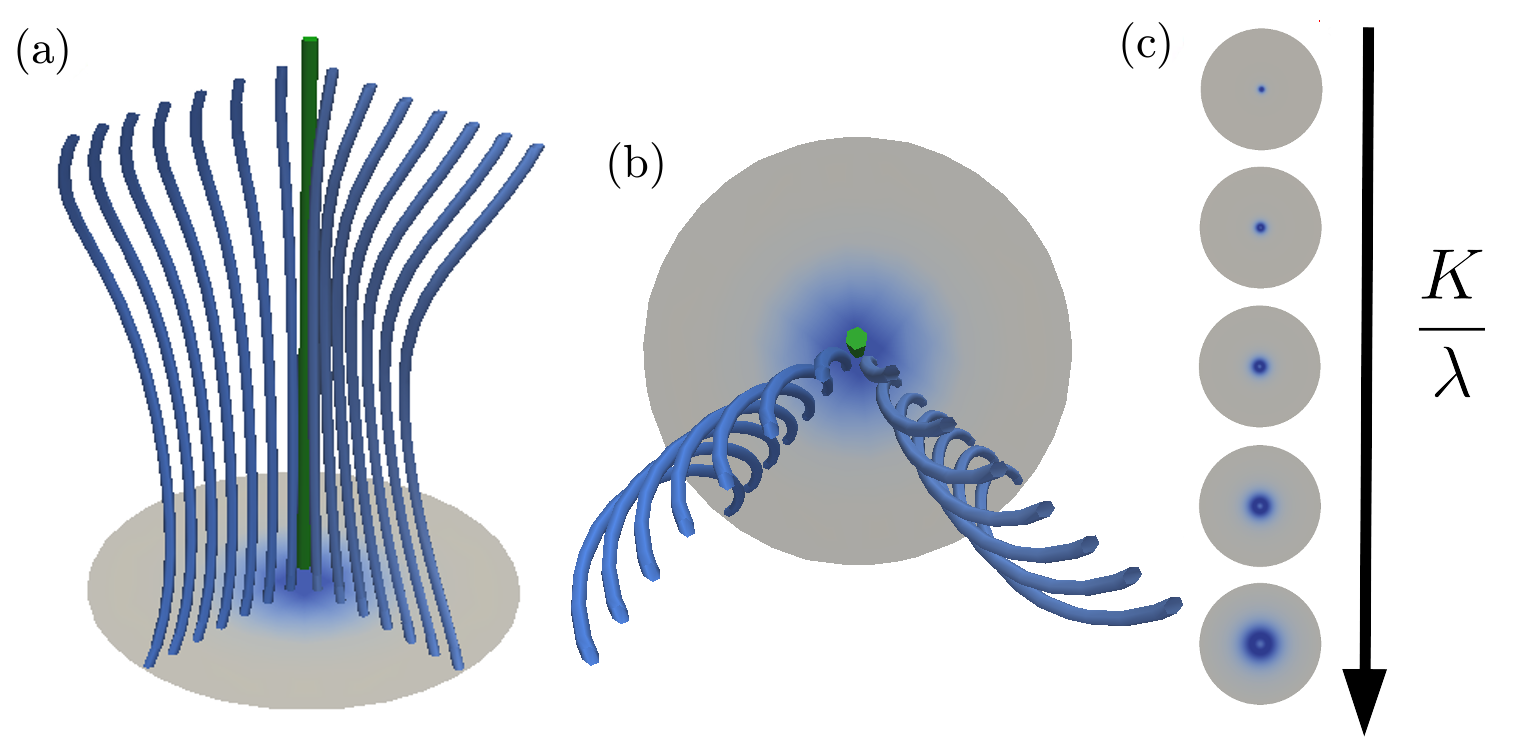}
  \caption{Illustration of the core structure of a screw dislocation. The core can be identified with the region where the cone angle $\theta$ deviates from the preferred cone angle $\theta_0$. (a,b) We plot the value of $|\theta-\theta_0|$ on a slice orthogonal to the $\beta$ line in a numerical simulation of a screw dislocation. The region where this is positive is shown in blue. The integral curves deform from helices to a straight line, where $\theta=0$, along the $\beta$ line. (c) The size of the core region is shown for several values of $K/\lambda$, which doubles with each panel, illustrating a roughly linear scaling.}
  \label{Sfig:core}
\end{figure}

As $\rho$ decreases and you approach the axis the integral curves are no longer exactly helices, however the $2\pi s$ winding of the bend vector is preserved. In \S\ref{sec:Local} we give a general analysis of the director structure as we approach a degenerate point and in \S\ref{sec:Meron} we describe some global, topological aspects. Here, we will continue to think of the integral curves as approximately helices but with curvature and torsion that vary with $\rho$, which is a good approximation provided the tilt angle $\theta_0$ is small. More precisely, consider the magnitudes of the two terms in \eqref{eq:Bend},
\begin{align}
  {\bf n} \cdot \nabla \theta & = \sin\theta \,\theta^{\prime}(\rho)\cos\bigl( qz+(s-1)\arctan(y/x) \bigr), \label{eq:term1} \\
\sin\theta({\bf n} \cdot \nabla \phi) & = \sin\theta \biggl( q\cos\theta + \frac{\sin\theta(\rho)}{\rho} \sin\bigl( qz+(s-1)\arctan(y/x) \bigr) \biggr).  \label{eq:term2}
\end{align}
Note that \eqref{eq:term2} shows that we require $\theta(\rho)$ to vanish at least linearly at the origin. The ratio of the two terms is then approximately $\theta^{\prime}(0) / (q + \theta^{\prime}(0))$ and taking $\theta^{\prime}(0)$ to be roughly $\theta_0$ divided by the pitch the ratio is of order $\theta_0 / 2\pi$ and is small. We can then neglect \eqref{eq:term1}, and simplify \eqref{eq:term2} to $|{\bf b}|=q\sin\theta(\rho) \cos\theta(\rho)$, the curvature of an integral helix. As $\rho \rightarrow 0$ this curvature vanishes, and along the $z$ axis itself the helices degenerate to a straight line, which is also our $\beta$ line. A schematic of this degeneration is shown in Fig.~\ref{fig:Screw}(a,b)(vi) and can be compared against numerical relaxation of a screw dislocation shown in Fig.~\ref{Sfig:core}. We identify the core region of the $\beta$ line by measuring how the cone angle $\theta$ deviates from the preferred value $\theta_0$ of the heliconical state and indicate it by blue shading. On the right, we show the size of the core region for different values of $K/\lambda$, corresponding to the helical pitch, increasing from top to bottom. The value of $K/\lambda$ doubles with each panel, illustrating a roughly linear scaling. The final panel is illustrated in more detail on the left of Fig.~\ref{Sfig:core}; compare with Fig.~\ref{fig:Screw}(a)(vi).

\subsection{Edge Dislocations}
\label{subsec:edge}

\begin{figure}[tbp]
\centering
\includegraphics[width=0.9\linewidth]{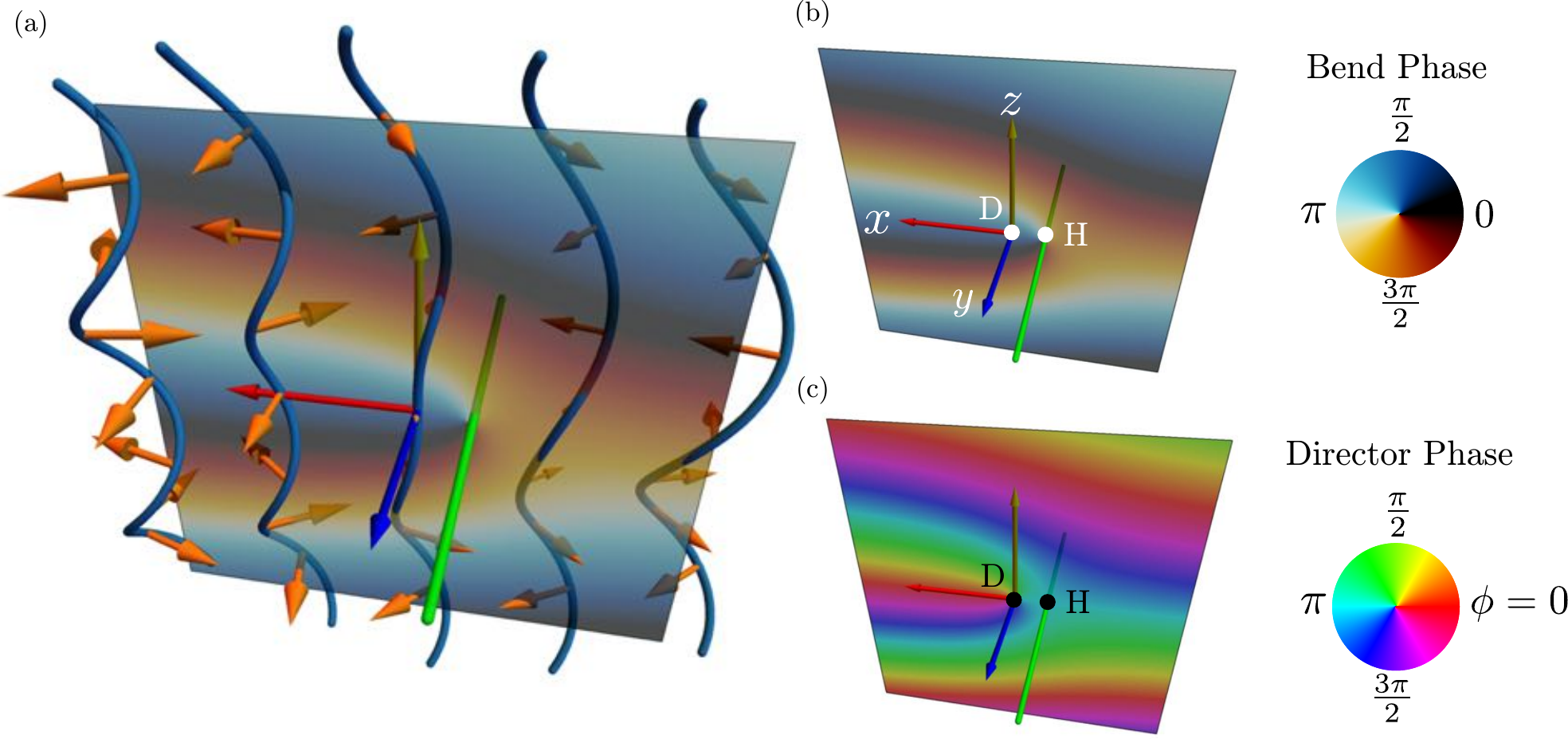}
    \caption{Edge dislocations in twist-bend nematics. (a) $xz$ slice through the edge dislocation parallel to the $y$ axis, coloured by the angle the bend vector makes with the $x$ axis, with director integral curves shown in blue, bend vector in orange and the $\beta$ line shown in green. Across the dislocation the bend acquires a $2 \pi$ winding. (b, c) The $\beta$ line does not coincide with the phase singularity $\bf D$ along the $y$ axis, but is along the hyperbolic line $\bf H$. We emphasise this difference by showing the angle the bend vector makes with the $x$ axis in (b), and the phase field $\phi$ in (c) --- note the discrepancy in the location of singularities.}
\label{fig:Edge}
\end{figure}

Returning to \eqref{eq:Screw} but taking instead $\phi = qz + s\arctan(z/x)$ yields an edge dislocation in the phase field parallel to the $y$ axis --- the case $s=+1$ is shown in Fig.~\ref{fig:Edge}. As we go from negative to positive $x$ an extra $2\pi$ is inserted into $\phi$, corresponding to an additional full turn in the integral helices, as can be seen in Fig.~\ref{fig:Edge}(a). On a positively oriented loop encircling the edge dislocation, the bend therefore acquires a winding of $2\pi s$ as in the case of the screw dislocation. There are, however, several distinct features of the edge dislocation worth emphasising. The first is that the $\beta$ line (shown in green in Fig.~\ref{fig:Edge}) is not itself an integral curve of the director --- this is the generic situation in an arbitrary director field, the screw dislocation being an exceptional case. The second feature is the location of the $\beta$ line itself --- it is not along the $y$ axis, but slightly displaced from it, as shown in Figs.~\ref{fig:Edge}(a,b). To understand this feature we recall some details of the phase field $\phi$, shown in Fig.~\ref{fig:Edge}(c) \cite{kamien2016}. An edge dislocation is composed of two disclinations in $\nabla\phi/|\nabla \phi|$. The first is a $+1$ disclination along the $y$ axis, denoted $\bf{D}$ in Figs.~\ref{fig:Edge}(b, c), which is a singularity in $\phi$. The second is a $-1$ disclination along $(-\frac{1}{q},y,0)$, called the hyperbolic line and denoted $\bf H$ in Figs.~\ref{fig:Edge}(b, c). This second disclination is the unique location where $\nabla\phi=0$, with $\phi$ itself nonsingular. We now return to \eqref{eq:Bend}, derived for the screw dislocation but valid here too. Neglecting $({\bf n} \cdot \nabla \theta)$ as before, we see $\bf b$ vanishes when $\nabla \phi$ vanishes, and so we have a $\beta$ line along the hyperbolic line $\bf H$. One might worry about the phase singularity at the origin, but a direct expansion of \eqref{eq:Bend} shows that the bend is in fact continuous about the origin, taking value ${\bf b} = \theta^{\prime}(0) \,{\bf e}_y$ at the origin itself, and is not (as one might initially suspect) singular --- this is reflected in the smooth nature of the bend at the origin shown in Figs.~\ref{fig:Edge}(a,b).

We briefly remark that the canonical local form of a family of curves which pass through an inflectional configuration (where the bend vanishes) is given in \citep{moffatt1992}, where it is shown that on passing through the inflectional configuration the curve normal (equivalently the bend $\bf b$) picks up a $2\pi$ rotation. Locally, this is what happens to our integral curves as we pass through the $\beta$ line at $\bf H$.

\subsection{Twist Grain Boundary Phases}
\label{sec:TGB}

The examples of screw and edge dislocations extend to constructions of locally heliconical director fields whose helical phase corresponds to any smectic texture. A general director field with these properties is given by
\begin{equation}
{\bf n} = \cos\theta \,{\bf N} + \sin\theta \bigl[ \cos\phi \,{\bf e}_1 + \sin\phi \,{\bf e}_2 \bigr] ,
\label{Seq:smectic_director}
\end{equation}
where $\phi$ is a smectic phase field, ${\bf N}$ is the smectic-A director field ({\sl i.e.} ${\bf N} = \nabla\phi / |\nabla\phi|$ away from singularities in $\phi$) and ${\bf e}_1$, ${\bf e}_2$ are an orthonormal basis for the planes orthogonal to ${\bf N}$ chosen to have no rotation along the integral curves of ${\bf N}$, meaning $(\nabla_{{\bf N}} {\bf e}_1) \cdot {\bf e}_2 = 0$. As in the screw and edge dislocation examples, the cone angle $\theta$ should vanish along the phase singularities. In this section and the next we outline constructions of this form for phase fields representing twist grain boundary and parabolic focal conic textures.

Twist grain boundaries in smectics are formed by arrays of equally spaced screw dislocations and mediate a rotation of the smectic layer normal. This same structure can be encoded into a director field that locally corresponds to the heliconical state; the grain boundary mediates a rotation of the helical (pitch) axis and each of the screw dislocations becomes a $\beta$ line. We first review briefly the construction of grain boundaries in smectics.

\begin{figure}[tb]
\centering
\includegraphics[width=\linewidth]{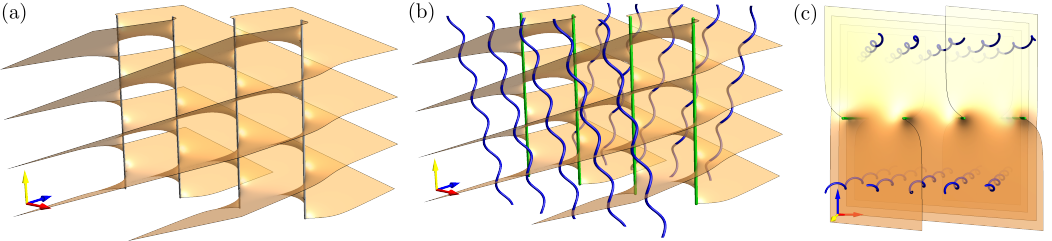}
    \caption{(a) Smectic phase field for a single grain boundary. The surface shown is $\phi = 0$, where $\phi$ is given in~\eqref{Seq:single_grain}. (b-c) Helical integral curves (blue) of a twist-bend director containing a grain boundary, with $\beta$ lines shown in green: (b) side view; (c) top view.}
\label{Sfig:TGB}
\end{figure}

A single grain boundary in a smectic can be described by the phase field~\cite{matsumoto2017}
\begin{equation}
\phi = \textrm{Im} \ln \Bigl[ \mathrm{e}^{-y/\ell} \mathrm{e}^{i \phi_{-}} + \mathrm{e}^{y/\ell} \mathrm{e}^{i \phi_{+}} \Bigr] ,
\label{Seq:single_grain}
\end{equation}
where $\phi_{\pm} = qz \cos(\alpha/2) \pm qx \sin(\alpha/2)$ and we choose $\ell = [q\sin(\alpha/2)]^{-1}$ to make $\phi$ a harmonic function. The layer structure is the level set $\phi=0$ and is shown in Fig.~\ref{Sfig:TGB}(a). For $y\lesssim -\ell$ we have $\phi \approx \phi_{-}$ and for $y\gtrsim \ell$ we have $\phi \approx \phi_{+}$. In the plane $y=0$ there are screw dislocations with axes parallel to $z$ at $x=\frac{\pi}{2} + m \pi$, $m\in\mathbb{Z}$. The gradient of the phase field is
\begin{equation}
\nabla\phi = q\cos(\alpha/2) \,{\bf e}_z + q \sin(\alpha/2) \,\frac{\sinh(2y/\ell) \,{\bf e}_x + \sin(2x/\ell) \,{\bf e}_y}{\cosh(2y/\ell)+\cos(2x/\ell)} ,
\end{equation}
and its magnitude squared,
\begin{equation}
|\nabla\phi|^2 = q^2 \,\frac{\cosh(2y/\ell)+\cos\alpha \cos(2x/\ell)}{\cosh(2y/\ell)+\cos(2x/\ell)} ,
\end{equation}
diverges as inverse distance squared along each of the screw dislocations.
It is not difficult to extend this construction to create phase fields containing multiple grains and describing full twist-grain boundary phases. We refer the reader to~\cite{matsumoto2017} for details.

We restrict our focus here to describing how the single grain boundary~\eqref{Seq:single_grain} can be embedded into a heliconical director field with $\beta$ lines along each of the screw dislocations, {\sl i.e.} the lines $(\frac{\pi}{2} + m\pi , 0 , z)$, $m\in\mathbb{Z}$. We write the director field in the form~\eqref{Seq:smectic_director} and take the basis $\{{\bf N}, {\bf e}_1, {\bf e}_2\}$ to be
\begin{align}
& {\bf N} = \cos\sigma \,{\bf e}_z + \sin\sigma \,{\bf e}_x , && {\bf e}_1 = - \sin\sigma \,{\bf e}_z + \cos\sigma \,{\bf e}_x , && {\bf e}_2 = {\bf e}_y ,
\end{align}
where $\sigma$ is a function interpolating between $-\alpha/2$ for $y \lesssim -\ell$ and $+\alpha/2$ for $y \gtrsim +\ell$, for instance $\sigma = \frac{\alpha}{2} \tanh(2y/\ell)$. With this choice ${\bf N}$ differs from $\nabla\phi/|\nabla\phi|$ by exponentially small terms away from the cores of the screw dislocations, along each of which it is ${\bf e}_z$. To make the cone angle $\theta$ vanish linearly along each screw dislocation and approach a preferred value $\theta_0$ outside of the core region we can choose $\theta = q\theta_0 / |\nabla\phi|$. A selection of helical integral curves of this director field are shown in Fig.~\ref{Sfig:TGB}(b,c).

\subsection{Parabolic Focal Conics}
\label{sec:FocalConic}

Focal conics are amongst the most celebrated geometric features of any ordered phase. They are the hallmark of smectic order, corresponding to the fundamental singularities of a material composed of equally spaced layers. They are also seen in twist-bend nematics~\cite{kleman2018}, which serve to emphasise that it is the one-dimensional periodicity that leads to focal conics, rather than a modulation of the mass density. A director field for a twist-bend phase containing a focal conic defect can be constructed using the general form~\eqref{Seq:smectic_director}, where $\phi$ is the phase field of a focal conic and ${\bf N}$ is the layer normal, away from the conic singularities themselves. The construction and description of the Dupin cyclides and focal conic domains is classical; here, we simply quote the formulae with a convenient parameterisation~\cite{alexander2010}.

\begin{figure}[tb]
\centering
\includegraphics[width=\linewidth]{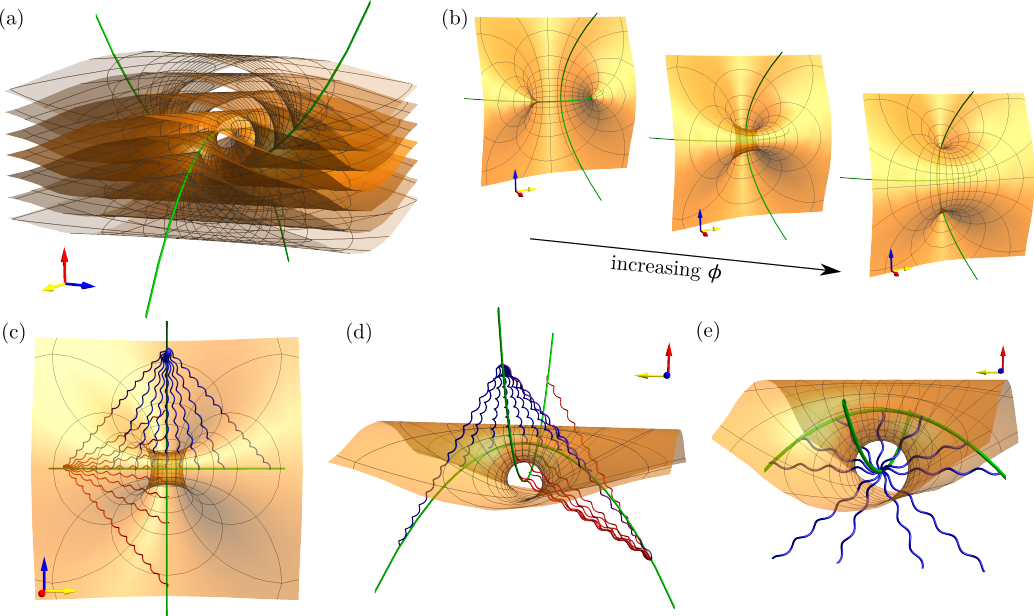}
    \caption{(a) Smectic phase field for a parabolic focal conic domain. We show multiple different level sets $\phi = \text{constant}$. The $\beta$ lines, singularities in the phase field $\phi$, are shown in green. (b) Individual layers in the parabolic focal conic domain are shown for increasing levels of $\phi$. (c-e) Helical integral curves of a twist-bend director containing a parabolic focal conic domain: (c) top view; (d) side view; (e) the local structure around each focus / apex is (compatible with) that of a (chiral) point defect. The integral curves connect one focus/ $\beta$ line to the other. We show two families of integral curves, one in red, one in blue, that converge on the same point on one of the foci.}
\label{Sfig:FocalConic}
\end{figure}

A focal conic domain consists of a space-filling family of surfaces -- level sets of a phase field $\phi$ -- that are singular along a pair of confocal conics and uniformly spaced everywhere else. In the case of a parabolic domain, the confocal parabolae may be taken to be
\begin{align}
& {\bf p}_1(u) = \biggl( \sigma \frac{\cos u}{1+\cos u} , \sqrt{2} \sigma \frac{\sin u}{1+\cos u} , 0 \biggr) ,
&& {\bf p}_2(v) = \biggl( -\sigma \frac{\cos v}{1+\cos v} , 0 , \sqrt{2} \sigma \frac{\sin v}{1+\cos v} \biggr) ,
\end{align}
where $\sigma$ is a constant parameter corresponding to the distance between the two foci/apices of the parabolae and $-\pi < u,v < \pi$. The domain itself then has the explicit parameterisation
\begin{equation}
\begin{split}
x & = \cos u \,\frac{\sigma - \phi (1+\cos v)}{2 + \cos u + \cos v} - \cos v \,\frac{\sigma + \phi (1+\cos u)}{2 + \cos u + \cos v} , \\
y & = \sqrt{2} \,\sin u \,\frac{\sigma - \phi (1+\cos v)}{2 + \cos u + \cos v} , \\
z & = \sqrt{2} \,\sin v \,\frac{\sigma + \phi (1+\cos u)}{2 + \cos u + \cos v} ,
\end{split}
\end{equation}
where each surface of constant $\phi$ is a parabolic Dupin cyclide.
Depending on the value of $\phi$ the range of $u,v$ should be restricted so as to terminate the surface on the singular parabolae. Specifically, if $\phi<-\sigma/2$ then the range of $u$ should be restricted according to $\cos u < |\sigma/\phi| - 1$; if $\phi>\sigma/2$ then the range of $v$ should be restricted by $\cos v < |\sigma/\phi| - 1$; and if $-\sigma/2 < \phi < \sigma/2$ no restriction is needed. In Fig.~\ref{Sfig:FocalConic}(a) we show the structure of a parabolic focal conic domain, with a selection of individual layers shown in Fig.~\ref{Sfig:FocalConic}(b).

In terms of this parameterisation the frame $\{{\bf N}, {\bf e}_1, {\bf e}_2\}$ is given by
\begin{equation}
\begin{split}
{\bf N} & = \biggl( - \,\frac{\cos u + \cos v + 2 \cos u \cos v}{2 + \cos u + \cos v} , - \,\frac{\sqrt{2} \sin u (1+\cos v)}{2 + \cos u + \cos v} , \frac{\sqrt{2} \sin v (1+\cos u)}{2 + \cos u + \cos v} \biggr) , \\
{\bf e}_1 & = \biggl( \frac{\sqrt{2} \sin u (1+\cos v)}{2 + \cos u + \cos v} , - \frac{1 + 2 \cos u + \cos u \cos v}{2 + \cos u + \cos v} , - \,\frac{\sin u \sin v}{2 + \cos u + \cos v} \biggr) , \\
{\bf e}_2 & = \biggl( \frac{\sqrt{2} \sin v (1+\cos u)}{2 + \cos u + \cos v} , \frac{\sin u \sin v}{2 + \cos u + \cos v} , \frac{1 + 2 \cos v + \cos u \cos v}{2 + \cos u + \cos v} \biggr) .
\end{split}
\end{equation}
Helical integral curves of the director field are then given by
\begin{equation}
{\bf h}_{(u,v)}(\phi) = {\bf x}_{0}(u,v) + \frac{\phi}{q} \,{\bf N} + \frac{\tan\theta}{q} \bigl[ \sin \phi \,{\bf e}_1 + \bigl( 1 - \cos \phi \bigr) {\bf e}_2 \bigr] ,
\end{equation}
where ${\bf x}_{0}(u,v)$ is a point on the cyclide $\phi = 0$. The range of values of $\phi$ should be limited to $[\frac{-\sigma}{1+\cos u}, \frac{\sigma}{1+\cos v}]$ and the helices then extend from one conic to the other. A selection of such helical integral curves are shown in Fig.~\ref{Sfig:FocalConic}(c-e). In this structure the two focal parabolae are singularities and correspond to $\beta$ lines. Although there are several possibilities for how the director is resolved along these lines, one natural arrangement places point defects at each focus/apex of the two parabolae; this local structure is especially suggested by Fig.~\ref{Sfig:FocalConic}(e).

\subsection{Skyrmions and Double Twist Cylinders}
\label{sec:Skyrmion}

\begin{figure}[tbp]
\centering
\includegraphics[width=0.99\linewidth]{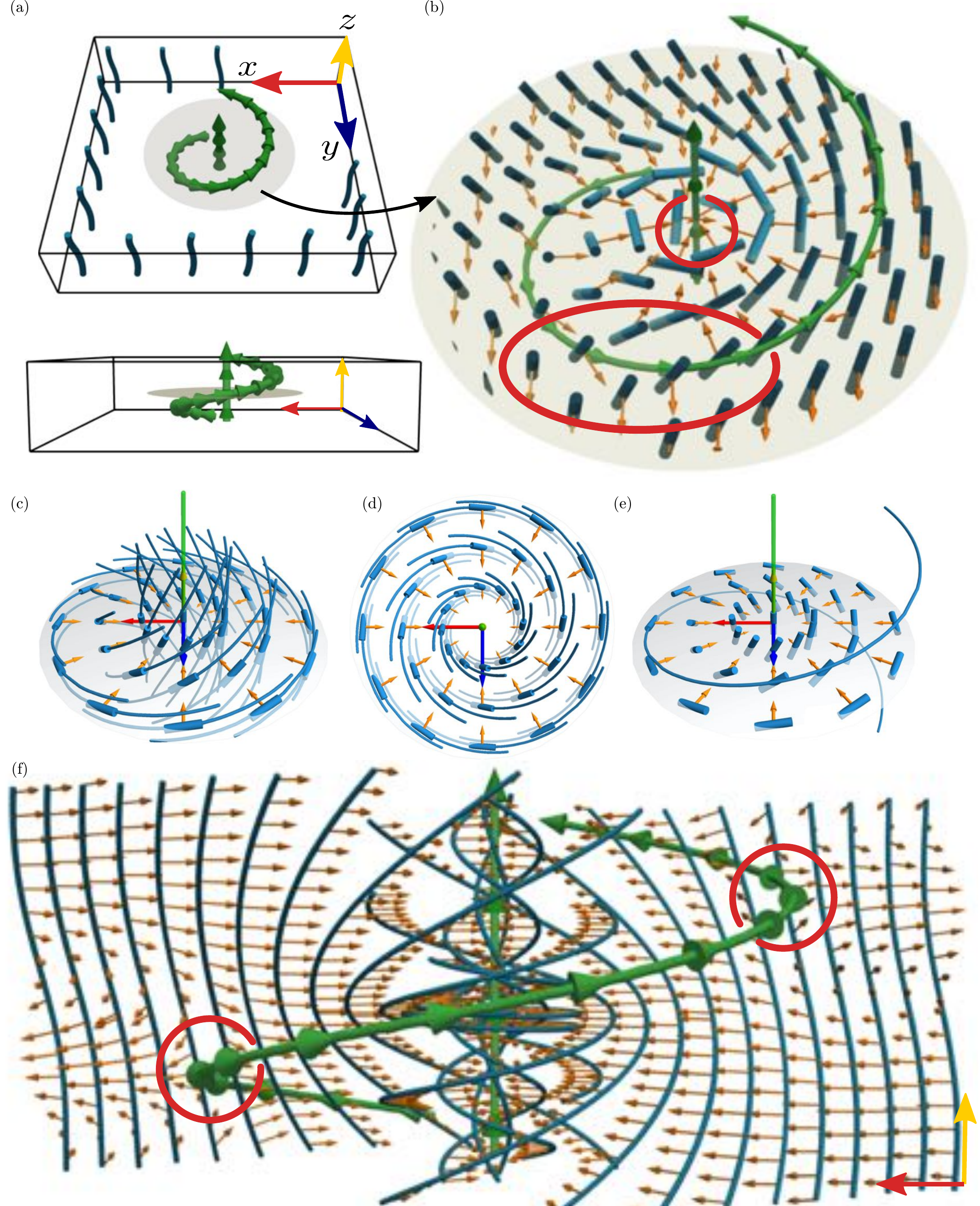}
    \caption{Twist-bend skyrmion in a heliconical background. (a) The two $\beta$ lines comprising the Skyrmion are shown in the simulation box, with grey disc in the midplane indicating rough Skyrmion extent. The background director is heliconical (blue curves). (b) Director and bend vector on midplane through the Skyrmion. Red circles highlight the winding of the bend around the $\beta$ lines. (c,d,e) Idealised double twist cylinder forming the neighbourhood of the vertical $\beta$ line. The director integral curves link the $\beta$ line, as emphasised in panel (e). (f) Director integral curves and bend vector on an $xz$ slice through the Skyrmion texture. Red circles emphasise the winding of the bend vector about the second helical $\beta$ line.}
\label{Sfig:Skyrmion}
\end{figure}

We now examine a class of defects in twist-bend nematics which are not constructed by analogy to a smectic phase field but rather from topologically non-trivial textures in cholesterics. Skyrmions are non-singular field configurations found in cholesterics and chiral ferromagnets~\cite{ackerman2014,ackerman2017,afghah2017,duzgun2018,foster2019,sutcliffe2017} corresponding to topologically protected particle-like solitons. They carry a topological charge $Q=\frac{1}{4\pi} \int {\bf n} \cdot \partial_x {\bf n} \times \partial_y {\bf n}\ dx dy$, an element of $\pi_2(S^2) \approx \mathbb{Z}$ giving the `wrapping number' of the texture. Given the general similarities between the heliconical director field and the cholesteric ground state it is natural to consider if Skyrmion textures also exist in twist-bend nematics and how they may be characterised in terms of $\beta$ lines and the geometry of bend.

In cholesterics, Skyrmions are usually created in frustrated cells with normal anchoring boundary conditions; away from the Skyrmion the director points vertically (say) so that the asymptotic behaviour is frustrated and not the cholesteric ground state. However, in twist-bend nematics the heliconical ground state may have a small cone angle (indeed arbitrarily small) allowing the usual Skyrmion structure to match naturally onto it as an asymptotic far field and this is the configuration we consider. In Fig.~\ref{Sfig:Skyrmion} we show a single Skyrmion embedded in a heliconical background, the result of numerical relaxation of \eqref{Seq:TB_energy} from a topologically correct initial director field. The Skyrmion is characterised by two $\beta$ lines as shown in Figs.~\ref{Sfig:Skyrmion}(a,b), the first vertical, the second a helix with pitch equal to the heliconical background. In the neighbourhood of the vertical $\beta$ line the director field is a double-twist cylinder, an idealised description of which is the texture ${\bf n} = \cos q\rho \,{\bf e}_z + \sin q\rho \,{\bf  e}_\phi$. In Figs.~\ref{Sfig:Skyrmion}(c,d,e) we show this texture, its integral curves and its bend vector. The texture has bend ${\bf b }= -\frac{1}{\rho} \sin^2 q\rho \,{\bf e}_{\rho}$, which vanishes linearly at the origin with winding number $+1$, giving a $\beta$ line along the $z$ axis. In contrast to the screw or edge dislocations discussed in \S\S\ref{subsec:screw},\ref{subsec:edge}, here the integral curves of the director link the $\beta$ line. This observation establishes that this $\beta$ line is topologically distinct from screw or edge dislocations, in the sense that a homotopy of the director between a double twist cylinder and an screw/edge dislocation  would necessarily introduce new $\beta$ lines (related observations of the failure of standard homotopy theory to deal with order parameters coupled to the director are given in \cite{beller2014} for the case of umbilic lines in cholesterics). The local structure about the second $\beta$ line is that of the edge dislocation, as can be seen in the integral curve structure shown in Fig.~\ref{Sfig:Skyrmion}(f).

In \S\ref{sec:Local} we will define a global orientation for $\beta$ lines --- this orientation is shown as arrows along the $\beta$ lines in Fig.~\ref{Sfig:Skyrmion}. We briefly note that the $\beta$ lines of Fig.~\ref{Sfig:Skyrmion} are both oriented along $+z$, and both puncture the grey disc shown in Figs~\ref{Sfig:Skyrmion}(a,b) in the same sense. The apparent difference in local winding of the bend vector between them is misleading, as the oriented plane on which one should measure winding makes a half turn, with director, between the two $\beta$ lines. With this orientation defined, in \S\ref{sec:FrenetSerret} we will apply the Gauss-Bonnet-Chern theorem to these Skyrmion textures to show that the two $\beta$ lines of Fig.~\ref{Sfig:Skyrmion} are topologically required.

\begin{figure}[tbp]
\centering
\includegraphics[width=0.8\linewidth]{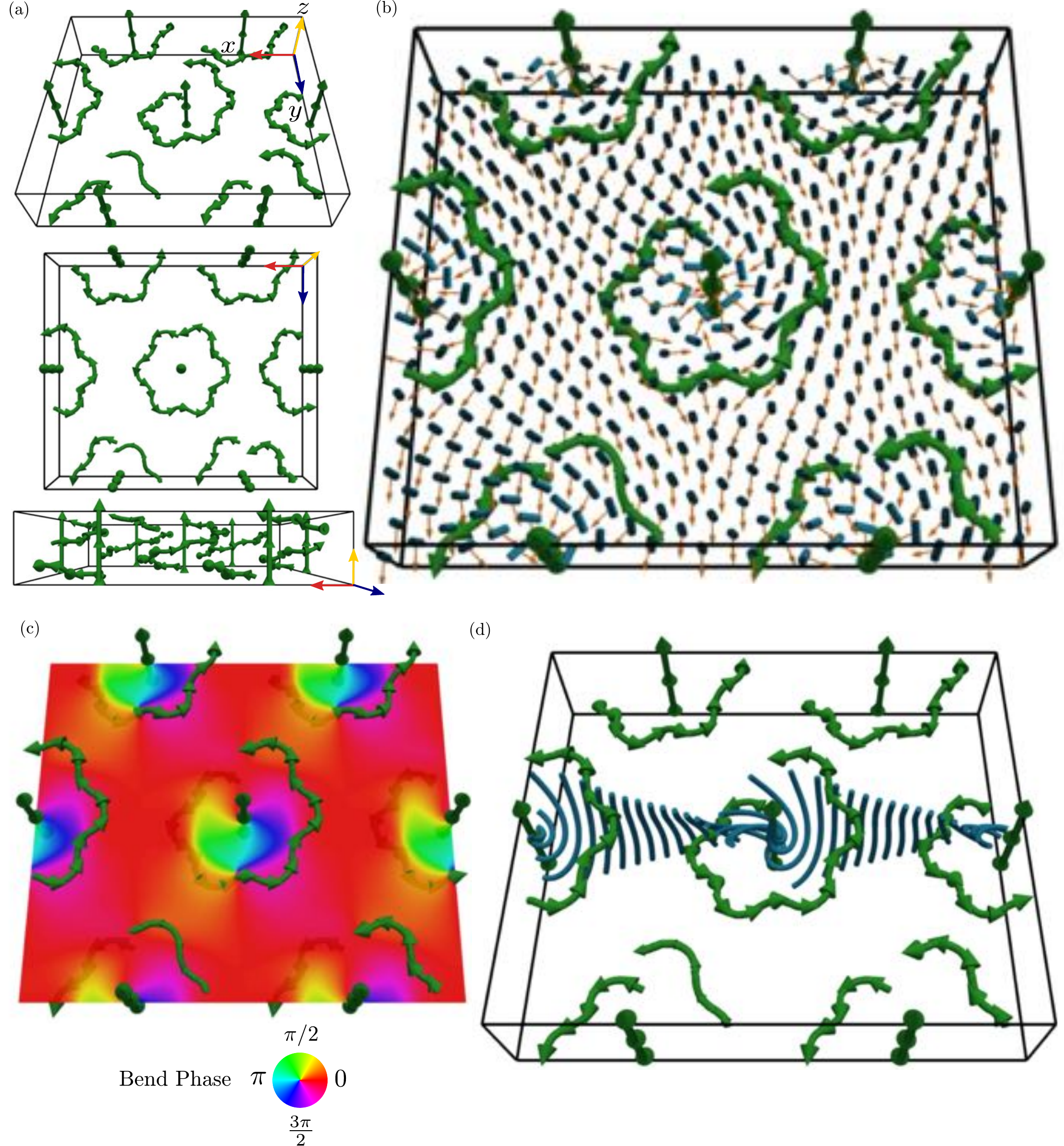}
    \caption{Skyrmion lattice in a twist-bend nematic. (a) Each Skyrmion in the lattice is composed of two $\beta$ lines, as in the isolated Skyrmion of Fig.~\ref{Sfig:Skyrmion}. The cylindrical symmetry of the helical $\beta$ line is broken to hexagonal by the lattice. (b) Director and bend vector on the midplane of the Skyrmion lattice --- compare with Fig.~\ref{Sfig:Skyrmion}(b). (c) Phase of the bend vector shown in panel (b). (d) Integral curves on an $xz$ slice through the lattice --- compare with Fig.~\ref{Sfig:Skyrmion}(f).}
\label{Sfig:SkyrmionLattice}
\end{figure}

In Fig.~\ref{Sfig:SkyrmionLattice} we show a lattice of Skyrmions, again obtained by numerical relaxation. The hexagonal symmetry of the lattice  breaks the cylindrical symmetry of the helical $\beta$ lines, but otherwise the texture is essentially that of a repeated isolated Skyrmion.

\section*{Local and Global Structure of $\beta$ Lines}

In the following sections we develop an account of the geometry and topology of the $\beta$ lines introduced in \S\ref{sec:general}. We discuss their local structure in \S\ref{sec:Local}, defining how $\beta$ lines may be oriented and showing how the director structure about the $\beta$ line sets its index. We then move to global structure in \S\ref{sec:FrenetSerret}, showing that $\beta$ lines are Poincar{\'e} dual to the Euler class of the plane field $\xi$, via an application of the Gauss-Bonnet-Chern theorem; concretely, these lines encode topological information about the director, such as Skyrmion number. We apply this general result to two specific examples, the screw dislocation \S\ref{subsec:screw} and an isolated Skyrmion \S\ref{sec:Skyrmion}.

\subsection{Local Analysis of Bend Zeros}
\label{sec:Local}

\begin{figure}[tb]
\centering
\includegraphics[width=0.5\linewidth]{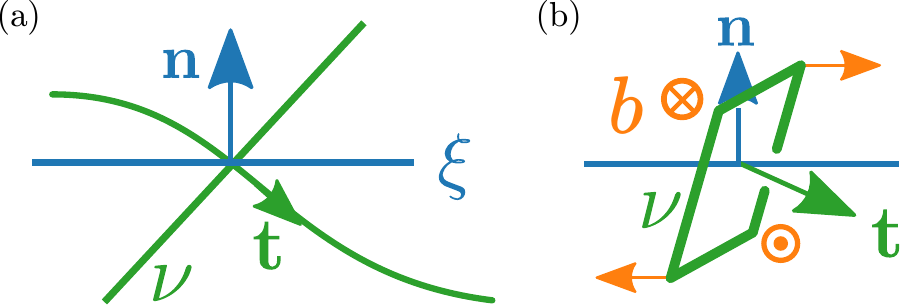}
  \caption{Local structure of a $\beta$ line. (a) $\beta$ lines come with two canonical frames, $\bf n$ and its orthogonal place $\xi$, and the tangent $\bf t$ and orthogonal plane $\nu$. (b) The bend vector locally lies in the plane $\xi$, making $\nabla \bf b|_\beta$ an isomorphism $\nu \rightarrow \xi$.}
\label{Sfig:Orientation}
\end{figure}

Along the $\beta$ line there are two canonical frames: One coming from the director and its orthogonal plane $\xi$, and the other coming from the tangent vector to the $\beta$ line ${\bf t}$ and its normal plane $\nu$ as shown in Fig.~\ref{Sfig:Orientation}(a). First we note that along the $\beta$ line the image of the linear map $\nabla {\bf b} |_{\beta} : T\mathbb{R}^3 \to T\mathbb{R}^3$ defined by ${\bf v} \mapsto ({\bf v}\cdot\nabla){\bf b}|_{\beta}$ is the orthogonal plane $\xi$, Fig.~\ref{Sfig:Orientation}(b). This is because ${\bf n}\cdot{\bf b}=0$ and hence $(\nabla{\bf b})\cdot{\bf n} = - (\nabla{\bf n})\cdot{\bf b}$, so that along a $\beta$ line $(\nabla{\bf b})\cdot{\bf n} = 0$. Similarly, the tangent vector ${\bf t}$ spans the kernel of $\nabla {\bf b} |_{\beta}$. This understood, we may think of $\nabla{\bf b}|_{\xi}$ as defining an isomorphism between the normal plane $\nu$ and the orthogonal plane $\xi$. The general linear group has two disconnected components, corresponding to positive and negative determinant. The orientation of the $\beta$ line is taken such that $\nabla{\bf b}|_{\xi}$ belongs to the positive, or orientation-preserving, component. In Figs.~\ref{Sfig:Skyrmion},\ref{Sfig:SkyrmionLattice} we indicate this orientation for the case of Skyrmions with arrows along the $\beta$ lines --- this orientation will enter into the signed intersection count with a surface which defines Skyrmion number in \S \ref{sec:FrenetSerret}. Note that under the replacement $\bf n \rightarrow -\bf n$, the bend $\bf b$ remains invariant, but the orientation of the plane field $\xi$ reverses, and hence all $\beta$ line orientations reverse. This reversal corresponds to the well-known reversal of Skyrmion number (hedgehog charge) under $\bf n \rightarrow -\bf n$~\cite{alexander2012}.

At any generic point the vectors ${\bf n}$ and ${\bf t}$ have no special relationship, being neither colinear nor perpendicular. Both these situations therefore correspond to situations of greater degeneracy. Points where ${\bf n}$ and ${\bf t}$ are perpendicular are the most basic type of degeneracy and have codimension one; we call them Legendrian points. Points of colinearity have codimension two. At a generic (or Legendrian) point, the planes $\nu$ and $\xi$ have one-dimensional intersection, which may be used to give a `framing', whose half-integer `self-linking' can change only by passing through points of colinearity.

We now relate $\nabla{\bf b}|_\beta$ to $\nabla \bf n$, computing the normal form of a Taylor series for the bend at a generic zero. The analysis closely parallels that for other geometric degeneracies such as umbilic points of surfaces~\cite{berry1977}, C lines in electromagnetic fields~\cite{nye1983} and umbilic lines in general~\cite{machon2016}. A Taylor series for a generic point where the bend vanishes will involve terms in the director field up to second order, so as to obtain all first order terms in the bend. Introducing a local coordinate system adapted to the director and its orthogonal plane at the bend zero, and writing ${\bf n} \approx n_x \,{\bf e}_x + n_y \,{\bf e}_y + {\bf e}_z$, we find the general form of the Taylor series contributing to the linear structure of the bend zero is
\begin{align}
\begin{bmatrix} n_x \\ n_y \end{bmatrix} & = \biggl[ \Bigl. \nabla_\perp {\bf n} \Bigr\rvert_0 + z \Bigl. \bigl( \partial_z\nabla_\perp {\bf n} \bigr) \Bigr\rvert_0 \biggr] \begin{bmatrix} x \\ y \end{bmatrix} + \frac{1}{2} z^2 \begin{bmatrix} s_x \\ s_y \end{bmatrix} ,
\label{Seq:director_local} \\
  \begin{bmatrix}b_x \\ b_y\end{bmatrix} &= \nabla {\bf b} \cdot  \begin{bmatrix} x \\ y \\z  \end{bmatrix} =  \biggl[ \Big( \Bigl. \nabla_\perp {\bf n} \Bigr\rvert_0 \Big)^2 + \Bigl. \partial_z\nabla_\perp {\bf n} \Bigr\rvert_0 \biggr] \begin{bmatrix} x \\ y \end{bmatrix} + z \begin{bmatrix} s_x \\ s_y \end{bmatrix} .
\label{Seq:bendprofile}
\end{align}
Here $\nabla_{\perp} {\bf n} = \Bigl[ \begin{smallmatrix} \partial_x n_x & \partial_y n_x \\ \partial_x n_y & \partial_y n_y \end{smallmatrix} \Bigr]$ denotes the 2$\times$2 matrix of orthogonal gradients of the director~\cite{machon2016} (see \S\ref{sec:general}), and $\partial_z \nabla_{\perp} {\bf n}$ is its rate of change along the local director. The winding number of the bend vector in the $xy$-plane is $\pm 1$ according to the sign of $\det\bigl( (\nabla_{\perp}{\bf n}|_{0})^2+\partial_z \nabla_{\perp}{\bf n}|_{0} \bigr)$; when the derivatives $\bigl. \partial_z \nabla_{\perp}{\bf n} \bigr|_{0}$ are negligible this reduces to $(\det \nabla_{\perp}{\bf n}|_{0})^2$ and the winding is always $+1$, so that the different profiles of $\beta$ lines are controlled crucially by the parallel derivatives of the orthogonal director gradients. $[s_x,s_y]$ controls the angle between the director and the tangent to the $\beta$ line. To see this, note that, as we saw above, \eqref{Seq:bendprofile} is a linear map $[x,y,z]\mapsto [b_x,b_y]$ with a one-dimensional kernel tangent to the $\beta$ line. When $[s_x,s_y]=0$ this kernel is along the $z$ axis.

\begin{figure}[tb]
\centering
\includegraphics[width=0.5\linewidth]{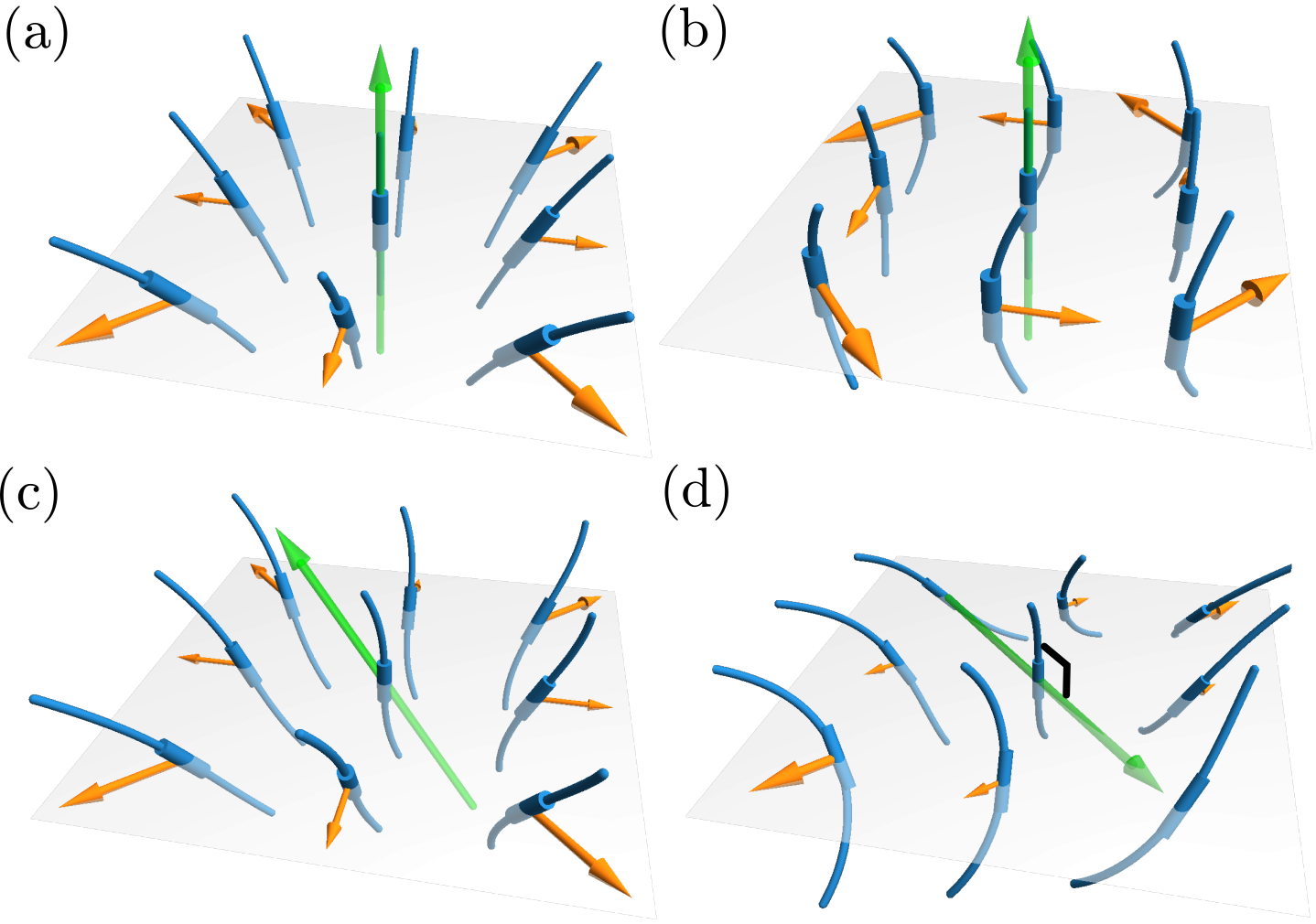}
  \caption{Local profiles of bend zeros from Taylor series, with director and its integral curves in blue, bend vector in orange and oriented $\beta$ line in green. (a) Radial $+1$ defect (b) azimuthal +1 defect. (c) Generic bend zero, with tilt between director and $\beta$ line. (d) Legendrian point, with degenerate winding behaviour.}
\label{Sfig:bend_profiles}
\end{figure}

With \eqref{Seq:bendprofile} we may construct $\beta$ lines with different local structures starting from a Taylor series for the director, with several examples shown in Fig.\ref{Sfig:bend_profiles}. In Fig.~\ref{Sfig:bend_profiles}(a), the only nonzero part of \eqref{Seq:bendprofile} is $(\nabla_\perp {\bf n})_{ij} = \delta_{ij} - n_i n_j$. This gives a pure splay distortion of the director, with a radial $+1$ defect in the bend along the $z$ axis. In Fig.~\ref{Sfig:bend_profiles}(b) we construct a vortex-like $+1$ defect by setting $(\partial_z \nabla_\perp {\bf n})_{ij} = \epsilon_{ij}$ with all else $0$. In Fig.~\ref{Sfig:bend_profiles}(c) we add a nonzero value of $[s_x,s_y]$ to the director field of Fig.~\ref{Sfig:bend_profiles}(a), which tilts the $\beta$ line. Finally, in Fig.~\ref{Sfig:bend_profiles}(d) we construct a Legendrian point where we encounter degenerate behaviour in the winding; this is done by arranging $\det\bigl( (\nabla_{\perp}{\bf n}|_{0})^2+\partial_z \nabla_{\perp}{\bf n}|_{0} \bigr)=0$.

\subsection{Frenet-Serret Frame, Connection and Curvature}
\label{sec:FrenetSerret}

At a generic point the director field carries a canonical Frenet-Serret framing. The Frenet-Serret frame associated to any space curve is the orthonormal frame consisting of its unit tangent, normal vector and binormal. As the bend is the curvature of the director integral curves its direction is exactly that of the Frenet-Serret normal for each integral curve. We write ${\bf b} = \kappa \,{\bf s}_1$, with $\kappa = |{\bf b}|$, and ${\bf s}_2 = {\bf n}\times{\bf s}_1$; the frame $\{{\bf n}, {\bf s}_1, {\bf s}_2\}$ gives a Frenet-Serret framing of the director field. It is defined on the complement of the $\beta$ lines, which are singularities of the Frenet-Serret framing.

The Frenet-Serret frame provides a canonical (Frenet-Serret) connection for the orthogonal plane field $\xi$
\begin{equation}
\omega = \bigl( \nabla {\bf s}_1 \bigr) \cdot {\bf s}_2 ,
\end{equation}
defined on the complement of the $\beta$ lines. The value of $\omega$ on the director field is the torsion, $\tau = \omega({\bf n}) = \bigl( \nabla_{{\bf n}} {\bf s}_1 \bigr) \cdot {\bf s}_2$, while the vector dual to it is the heliconical pitch axis; both are singular along the $\beta$ lines. The associated curvature 2-form (the curvature of the plane field $\xi$) is
\begin{equation}
\Omega = d\omega = \frac{-1}{2} \epsilon_{ijk} \,n_{i} \,dn_{j} \wedge dn_{k} .
\end{equation}
In the heliconical ground state we have $\omega = q \cos\theta \,dz$, the torsion is $\tau = q \cos^2\theta$, the heliconical pitch axis is ${\bf e}_z$ and the curvature vanishes. When the local helical structure varies slowly as in the director ${\bf n} = \cos\theta \,{\bf e}_z + \sin\theta [ \cos\phi \,{\bf e}_x + \sin\phi \,{\bf e}_y ]$ the Frenet-Serret connection is $\omega \approx \cos\theta \,d\phi$, the torsion is $\tau \approx \cos^2\theta \,|\nabla\phi|$, the heliconical pitch axis is $\nabla\phi / |\nabla\phi| = {\bf N}$ and the curvature is $\Omega = - \sin\theta \,d\theta \wedge d\phi$.

We remark that the bend is invariant under the nematic symmetry ${\bf n} \to -{\bf n}$ and as a consequence both the curvature $\kappa$ and Frenet-Serret normal ${\bf s}_1$ are unchanged under this transformation. On the other hand, the binormal ${\bf s}_2 = {\bf n} \times {\bf s}_1$ changes sign, as does the Frenet-Serret connection $\omega$ and curvature $\Omega$. This latter is the well-known change in sign of nematic hedgehog charge under ${\bf n} \to -{\bf n}$~\cite{alexander2012}. The heliconical pitch axis also reverses but the torsion $\tau$ is invariant.
Along a $\beta$ line the Frenet-Serret connection degenerates as a multiple of the angular form winding around it, which provides an orientation of the $\beta$ line; like the connection, this orientation reverses under ${\bf n} \to -{\bf n}$.

The integral of the curvature over a surface $S$ detects topological properties of the director field as described by the Gauss-Bonnet-Chern theorem
\begin{equation}
\frac{1}{2\pi} \int_{\partial S} \omega - \frac{1}{2\pi} \int_{S} \Omega = e_{\xi}(S) = \sum_{j} \textrm{Int}\bigl( \beta_{j} , S \bigr) ,
\label{Seq:GBC}
\end{equation}
where the Euler number $e_{\xi}(S)$ of the plane field $\xi$ can equally be calculated as the total intersection number of the surface with the $\beta$ lines by Poincar\'e duality. This number depends on the homology class of the surface $S$ relative to its boundary.
As an example, consider the screw dislocation textures of \S\ref{subsec:screw} and let $S$ be a disc of (large) radius $R$ in the plane $z=0$, centred on the origin. On the boundary of the disc where the director is locally the heliconical state with preferred cone angle $\theta_0$, the Frenet-Serret connection is $\omega = \cos\theta_0 \,d\phi$ with $\phi = qz + s \arctan(y/x)$ and
\begin{equation}
\frac{1}{2\pi} \int_{\partial S} \omega = \frac{\cos\theta_0}{2\pi} \int_{\partial S} \biggl( q \,dz + s \frac{-y \,dx + x \,dy}{x^2+y^2} \biggr) = s \cos\theta_0 .
\end{equation}
The curvature is $\Omega = - \sin\theta \,d\theta\wedge d\phi$ and its integral is (minus) the area swept out by the director field over $S$
\begin{equation}
\frac{1}{2\pi} \int_{S} \Omega = s \bigl( \cos\theta_0 - 1 \bigr) ,
\end{equation}
so that the Gauss-Bonnet-Chern theorem gives
\begin{equation}
\frac{1}{2\pi} \int_{\partial S} \omega - \frac{1}{2\pi} \int_{S} \Omega = s \cos\theta_0 - s \bigl( \cos\theta_0 - 1 \bigr) = s .
\end{equation}
The Euler number is the strength of the screw dislocation; as there is a single $\beta$ line along the $z$-axis it is also the intersection number of the $\beta$ line with $S$.

A similar example can be given for the double twist director of \S\ref{sec:Skyrmion}
\begin{equation}
{\bf n} = \cos\theta \,{\bf e}_z + \sin\theta \bigl[ \sin \arctan(y/x) \,{\bf e}_x - \cos \arctan(y/x) \,{\bf e}_y \bigr] ,
\end{equation}
that describes the core region of a Skyrmion. Here $\theta = \theta(\rho)$ is a function of the radial distance $\rho=\sqrt{x^2+y^2}$ from the axis of the cylinder, along which $\theta$ vanishes. The bend is
\begin{equation}
{\bf b} = \frac{-\sin^2\theta}{\rho} \bigl[ \cos \arctan(y/x) \,{\bf e}_x + \sin \arctan(y/x) \,{\bf e}_y \bigr] ,
\end{equation}
and the Frenet-Serret connection and curvature are
\begin{align}
& \omega = \cos\theta \frac{-y \,dx + x \,dy}{x^2+y^2} , && \Omega = \frac{-\sin\theta \,\theta^{\prime}}{\rho} \,dx\wedge dy .
\end{align}
Integrating over a disc of radius $R$ in the plane $z=0$ (centred on the axis) we have
\begin{equation}
\frac{1}{2\pi} \int_{\partial S} \omega - \frac{1}{2\pi} \int_{S} \Omega = \cos\theta(R) - \bigl( \cos\theta(R) - 1 \bigr) = 1 ,
\end{equation}
corresponding to the intersection number of the $\beta$ line along the axis of the double twist cylinder with the disc.
In a Skyrmion this core region of double twist connects smoothly to an asymptotic director corresponding to a pure heliconical ground state. As the bend vector of the double twist region has winding number $+1$ in the $xy$-plane, while that of the heliconical ground state is constant, there is necessarily a $\beta$ line involved in any such interpolation. For any surface $S$ extending into the heliconical ground state we have $\omega|_{\partial S} = q \cos\theta \,dz$ and
\begin{equation}
\frac{1}{2\pi} \int_{\partial S} \omega - \frac{1}{2\pi} \int_{S} \Omega = 0 - \Bigl[ \bigl( \cos\theta(R) - 1 \bigr) + \bigl( -1 - \cos\theta(R) \bigr) \Bigr] = 2 ,
\end{equation}
the additional contribution from the integrated curvature being the area (divided by $2\pi$) swept out on the unit sphere by the director field in the interpolation between the inner double twist region and asymptotic heliconical texture. Again, the Euler number is the total intersection number of $S$ with the $\beta$ lines, and is twice the Skyrmion charge.

\begin{figure}[tb]
\centering
\includegraphics[width=0.7\linewidth]{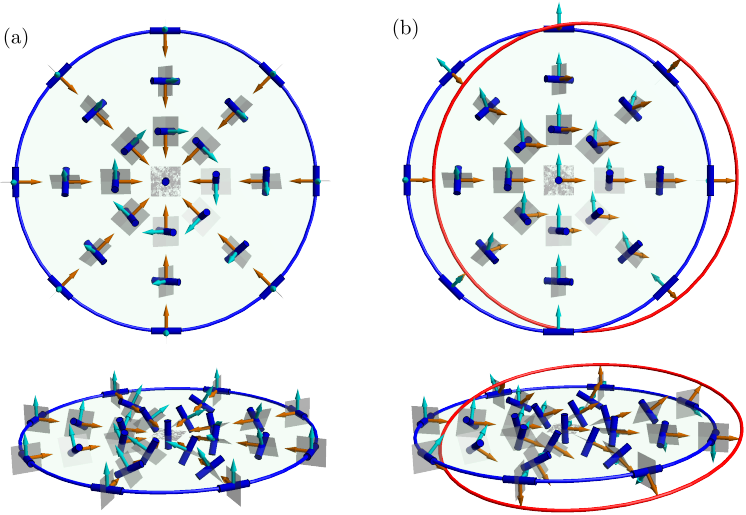}
\caption{Illustration of the transverse self-linking number associated to a closed integral curve of the director field, here a planar circle (blue). (a) The Frenet-Serret framing with the normal (direction of the bend) in orange and the binormal in cyan. The self-linking number of the closed integral curve with this framing is zero but the frame has a singularity at the centre of the disc. (b) A trivialisation of the orthogonal plane field over the disc induces a framing of the closed integral curve with self-linking $-1$ (illustrated using the red curve displaced from $K$ along a basis vector of the trivialisation); this is the transverse self-linking number of the curve. In both panels we show a top down view and a side-on view for clarity.}
\label{Sfig:SL}
\end{figure}

We finish this section by noting a connection to the C\u{a}lug\u{a}reanu theorem~\cite{calugareanu1961,fuller1971} that arises for closed integral curves of the director. Suppose $K$ is such a closed integral curve. Since the director is the tangent vector to this curve and $\omega({\bf n})=\tau$, the integral of the connection yields the twist of $K$, with its Frenet-Serret framing
\begin{equation}
\frac{1}{2\pi} \int_{K} \omega = \frac{1}{2\pi} \int_{K} \tau \,ds = \textrm{Tw}(K) .
\end{equation}
If $S$ is any Seifert surface for $K$ then the intersection number of the $\beta$ lines with $S$ is equal to the difference between the Frenet-Serret self-linking number of $K$, $\textrm{SL}(K)$, and the self-linking number of a framing that extends over $S$ without any singularities, which we call the transverse self-linking number, $\overline{\textrm{SL}}(K;S)$, a quantity of significance in contact topology~\cite{geiges2008,machon2016}. The transverse self-linking number is illustrated in Fig.~\ref{Sfig:SL} for the simplest example of a planar circle bounding a disc. With these two identifications~\eqref{Seq:GBC} becomes
\begin{equation}
\textrm{Tw}(K) - \frac{1}{2\pi} \int_{S} \Omega = \textrm{SL}(K) - \overline{\textrm{SL}}(K;S) .
\end{equation}
Finally, using the C\u{a}lug\u{a}reanu theorem~\cite{calugareanu61,fuller71}, $\textrm{SL}(K)=\textrm{Tw}(K)+\textrm{Wr}(K)$, where $\textrm{Wr}(K)$ is the writhe of $K$, we obtain a geometric integral formula for the transverse self-linking number
\begin{equation}
\overline{\textrm{SL}}(K;S) = \frac{1}{2\pi} \int_{S} \Omega + \textrm{Wr}(K),
\end{equation}
as a sum of the total Berry curvature of the Seifert surface and the writhe of the closed integral curve. Of course, the integrated curvature has the interpretation as the twist of $K$ with the transverse framing.

\subsection{Knots, Merons, Linking and Self-Linking}
\label{sec:Meron}

\begin{figure}[t]
\centering
\includegraphics[width=\linewidth]{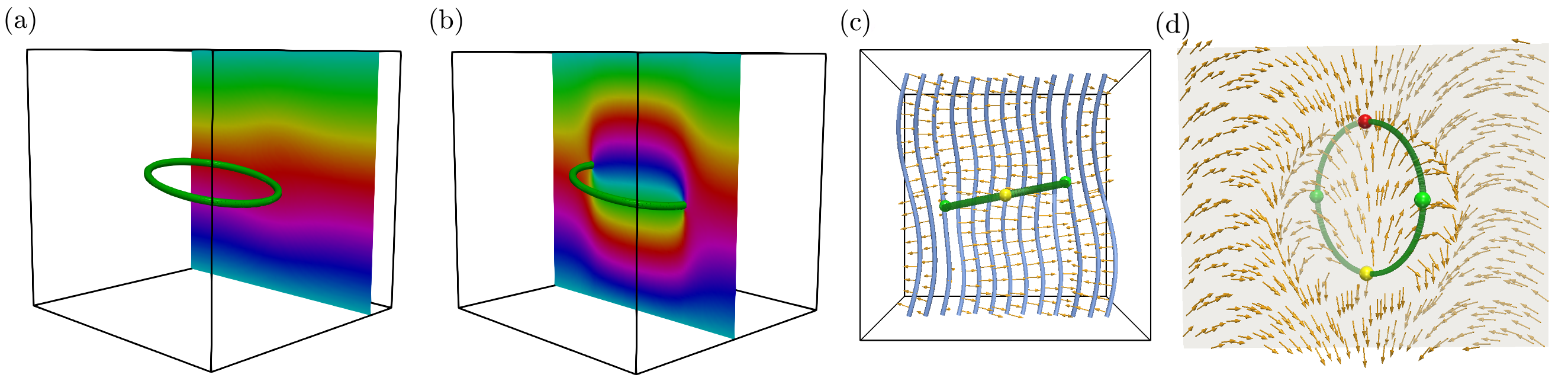}
\caption{A circular edge dislocation embedded in a heliconical background. The phase field $\phi$ is shown on (a) a surface away from the $\beta$ line (green), and (b) a surface that intersects the $\beta$ line. (c) There are two Legendrian points on the line, indicated by blue spheres. The director (blue curves) and bend (orange) are shown on a slice intersecting the two Legendrian points. (d) The bend is shown on a slice though the $\beta$ line, which intersects the $\beta$ line at points halfway between the Legendrian points, indicated by coloured sphere. At the purple point, the winding of the bend around the $\beta$ line is $-1$. At the yellow point, the winding of the bend around the $\beta$ line is $+1$, as can be seen from examing the bend vector field itself.}
\label{fig:edge_circle}
\end{figure}

In this final section we consider some examples of the global properties of $\beta$ lines, when they form closed loops, knots and links. These are relevant to the increasing number of complex, three-dimensional knotted fields~\cite{chen2013prl,ackerman2017prx,machon2016,machon2016prsa,sutcliffe2018,tai2019}, whose intricate structures realise knotted field lines, disclinations and geometric degeneracies, including umbilic and $\beta$ lines.
The simplest example is obtained by wrapping the edge dislocation discussed in \S\ref{subsec:edge} around an axis to form a circular loop. This example is shown in Fig.~\ref{fig:edge_circle} and illustrates several concepts from the preceding sections. First, we observe that the profile of the bend around the circular $\beta$ line changes as we move along it, from a $-1$ winding to a $+1$ winding. Consequently, there must be a pair of Legendrian points on the $\beta$ line. The local structure of the Legendrian points is given by the saddle-node bifurcation, where the winding around a critical point changes sign, as described in~\cite{etnyre1999}. If the $\beta$ line were flat, laying entriely in the $z$ plane, then every point would be Legendrian. This is non-generic, so the $\beta$ line is tilted out of this plane. The Legendrian points are indicated by blue spheres in Fig.~\ref{fig:edge_circle}(c,d). In Fig.~\ref{fig:edge_circle}(d), we show the bend on a slice that intersects the $\beta$ line at two points directly in between the Legendrian points, indicated by a yellow sphere, where the bend has winding $+1$ around the line, and a purple sphere, where the bend has winding $-1$.

As well as realising the unknot as a $\beta$ line, it is possible to embed an arbitrary knotted or linked set of $\beta$ lines into a heliconical background, via an extension of our constructions for screw and edge dislocations. Given any knot or link $K$, the director
\begin{equation}
{\bf n} = \cos\theta \,{\bf e}_z + \sin\theta \bigl[ \cos\phi_K \,{\bf e}_x + \sin\phi_K \,{\bf e}_y \bigr] ,
\label{eq:meron}
\end{equation}
where $\phi_{K} = qz + \frac{1}{2} \omega_{K}$, with $\omega_K$ the solid angle function for $K$~\cite{binysh2018}, embeds a helical winding of the director integral curves around a tubular neighbourhood of $K$; as before, the cone angle $\theta$ should be made to vary from its far field value to vanish along $K$. The phase winding in the helical integral curves guarantees the existence of a $\beta$ line.

\begin{figure}[t]
\centering
\includegraphics[width=\linewidth]{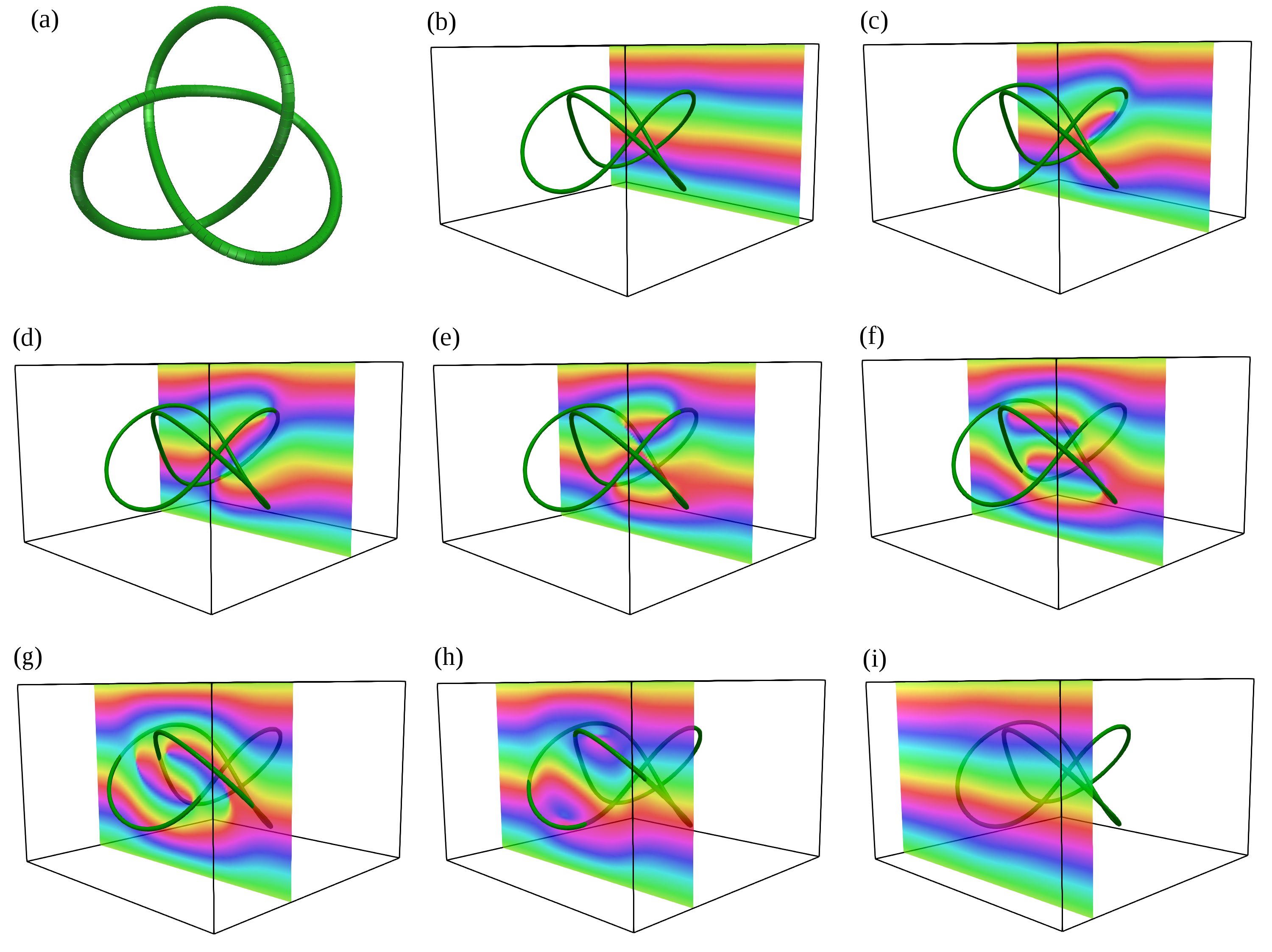}
\caption{A meron tube along a trefoil knot. (a) The $\beta$ line (green), shown from above. (b-i) The helical phase field on different slices through the texture. The meron tube is an edge dislocation, and the changes in the helical phase field shown on the slices as one passes through the $\beta$ line should be compared with the edge dislocation shown in Fig.~\ref{fig:Edge}.}
\label{fig:meron3D}
\end{figure}

\begin{figure}[t]
\centering
\includegraphics[width=\linewidth]{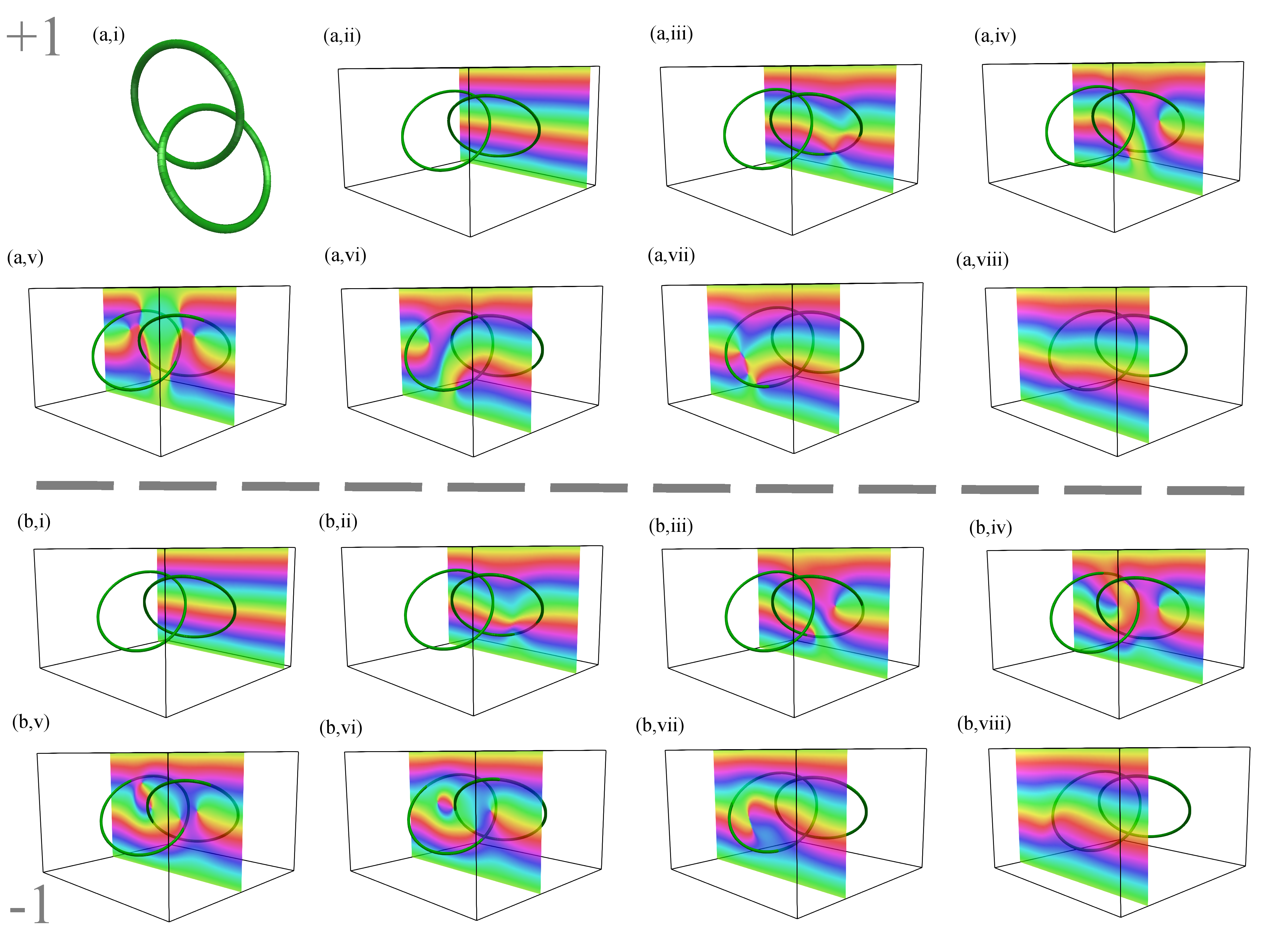}
\caption{A meron tube along (a) a Hopf link with linking number $+1$, and (b) a Hopf link with linking number $-1$. In each panel the $\beta$ lines are shown in green, and the colours on each slice show the helical phase field.}
\label{fig:hopf3D}
\end{figure}

The director texture is that of a meron tube extruded along $K$. A meron is a fractionalisation of a Skyrmion that carries half the topological charge~\cite{duzgun2018,yu2018}. $\beta$ lines provide a natural geometric perspective on this fractionalisation: since each Skyrmion comprises two $\beta$ lines, a single $\beta$ line represents half a Skyrmion, {\sl i.e.} a meron. In terms of the heliconical phase field, $\phi_{K}$, these meron tubes are edge dislocations where heliconical layers terminate. Exactly these structures were recently created experimentally in cholesteric cells and shown to form highly controllable and responsive knotted solitons~\cite{tai2019}. In that experiment, links of `escape up' and `escape down' meron tubes combined to give non-zero Hopf invariant. For the twist-bend nematic phase, the small conical angle ($\theta \approx 25^{\circ}$~\cite{chen2013}) gives an energetic preference to `escape up' merons over `escape down', whereas in cholesterics ($\theta = \pi/2$) the two types of meron are degenerate. An example for the trefoil knot is shown in Fig.~\ref{fig:meron3D}. The phase field $\phi$ is shown on several slices through the $\beta$ line, which is shown as a green curve in each panel. These slices should be compared with the structure of the phase field for an edge dislocation in Fig.~\ref{fig:Edge}. Panels (b-i) show the change in the phase field on a surface as one slides that surface across the $\beta$ line. The change in the number of layers as the surface crosses the $\beta$ line is clear from an examination of the phase field. Similar images are shown in Fig.~\ref{fig:hopf3D} for the two Hopf links, with linking number $+1$, Fig.~\ref{fig:hopf3D}(a), and $-1$, Fig~\ref{fig:hopf3D}(b).

Knotted meron tubes illustrate a further property of the $\beta$ lines, which capture not only the Euler class of the director and the Skyrmion charge (via the Gauss-Bonnet-Chern theorem \eqref{Seq:GBC}) but also the same information as the Hopf invariant. Classicially, three-dimensional knotted solitons in $S^3$ (or $\mathbb{R}^3$ with a uniform background director) are characterised by the homotopy group $\pi_3(S^2)$. The Hopf invariant establishes an isomorphism between this group and the integers, $\pi_3(S^2) \cong \mathbb{Z}$, and is computed via the linking of preimages. Gompf and Stipsicz~\cite{Gompf} offer an alternative way of describing this invariant which connects it to the zeros of a vector field orthogonal to the director, such as the bend. The invariant is a linking number,
\begin{equation} \label{eq:hopf}
  \Theta = \sum_{i} s_i^2 \mathrm{SL} ( \beta_i ) + \sum_{i\neq j} s_i s_j\mathrm{Lk} ( \beta_i , \beta_j ),
\end{equation}
familiar from helicity and abelian Chern-Simons theory~\cite{ArnoldKhesin}, where the $j$th $\beta$ line $\beta_j$ has strength $s_j$. The self-linking number, $\mathrm{SL}(\beta)$, is defined as follows: consider the total rotation $\int_{B^{\prime}} {\bf e}_2 \cdot d{\bf e}_1$ of the Frenet-Serret frame about the director along any push-off $B^{\prime}$ giving a zero-framing for the $\beta$ line. Part of this rotation is an intrinsic Berry phase $\gamma$, equal to the area on the unit sphere bound by the curve traced out by {\bf n} along $B$. The difference $\gamma - \int_{B^{\prime}} {\bf e}_2 \cdot d{\bf e}_1 = 2\pi \, \textrm{SL}(\beta)$ defines the self-linking. In the (non-generic) case where the pushoff $B^\prime$ is transverse to the planes $\xi$ orthogonal to the director, then the self-linking number just defined is the same as the self-linking number of $B^\prime$ computed by pushing off along the bend vector field. In general, there is no direct relationship between $\Theta$ and the Hopf invariant, however they capture the same fundamental topology, and a uniform state with vanishing Hopf invariant will also have vanishing $\Theta$.

For example, we may produce a director with Hopf invariant $H$ by taking a double-twist cylinder (Skyrmion tube) and twisting it $H$ times before joining the endpoints. The resulting solid torus can be embedded into a uniform background to give an `axially-symmetric' Hopfion~\cite{sutcliffe2018}. As we have disucssed, there are two $\beta$ lines, a central line with strength $+1$ and a second $\beta$ line wrapping around it with strength $-1$, as shown in Fig.~\ref{Sfig:Skyrmion}. Both lines have vanishing self-linking number, while the linking number of the two $\beta$ lines is $-1$, so that $\Theta = 2H$. As a second example, consider the trefoil knot shown in Fig.~\ref{fig:meron3D}. There is a single $\beta$ line corresponding to the green curve, and consequently the invariant $\Theta$ is equal to the self-linking number. In the construction we have given the framing on the $\beta$ line is the solid angle framing, so that the self-linking number vanishes, and also $\Theta=0$.

\end{document}